\documentclass[twocolumn,superscriptaddress,longbibliography,aps,prb,amsmath,preprintnumbers]{revtex4-1}
\usepackage{tikz}
\usepackage{hyperref}
\usepackage{epsfig}

\usepackage{graphicx}

\begin{document}

\title[Time-dependent Andreev bound states]{Time-dependent Andreev bound states of a quantum dot\\ 
        coupled to two superconducting leads}

\author{R. Taranko}
\author{T. Kwapi\'nski }
\author{T.\ Doma\'nski}
\email{doman@kft.umcs.lublin.pl}
\affiliation{Institute of Physics, M.\ Curie Sk\l odowska University, 20-031 Lublin, Poland}

\date{\today}
%
%
\begin{abstract}
Sub-gap transport properties of a quantum dot (QD) coupled to two superconducting and one metallic leads
are studied theoretically, solving the time-dependent equation of motion by the Laplace transform technique. 
We focus on time-dependent response of the system induced by a sudden switching on the QD-leads couplings, 
studying the influence of initial conditions on the transient currents and the differential conductance.
We derive analytical expressions for measurable quantities and find that they oscillate in time with 
the frequency governed by the QD-superconducting lead coupling and acquire damping,  due to relaxation 
driven by the normal lead. Period of these oscillations increases with the superconducting phase difference 
$\phi$. In particular, for $\phi=\pi$ the QD occupancy and the normal current evolve monotonically 
(without any oscillations) to their stationary values. In such case the induced electron pairing 
vanishes and the superconducting current is completely blocked. We also analyze time-dependent 
development of the Andreev bound states. We show, that the measurable conductance peaks do not 
appear immediately after sudden switching of the QD coupling to external leads but it takes 
some finite time-interval for the system needs create these Andreev states. Such time-delay 
is mainly controlled by the QD-normal lead coupling.
\end{abstract}

\maketitle

\section{Introduction}

Transient effects of the quantum dot (QD) systems have been intensively studied over last years, 
providing useful insight into the electron transport properties. These effects could be of special 
importance in experiments on nanoscopic devices, where different types of time-dependent pulses can
effectively control the electron flow. Transient effects have been studied, both theoretically and 
experimentally for the QDs coupled to the metallic (conducting) electrodes
\cite{001,002,003,004,005,006,007,008,009,010,011,012,013,014,015,016,017,018,019,020,021,022,023,024,025,026,027,028,029,030,031,032,033,034,035,036,037,038,039,040,041,042,043,044,045,046,047,048,049,050,051,052,053,054,055,056,056,058,059,060,061,062,063,064,065,066,067,068}
and in the presence of superconducting reservoirs
\cite{069,069,070,071,072,073,074,075,076,077,078,079,080,081,082,083,084,085}. Numerous theoretical
approaches have been developed to deal with such time-dependent problems, e.g. the iterative 
influence-functional path integral, \cite{035} Keldysh formalism and time-dependent partition-free 
approach, \cite{040} weak-coupling continuous-time Monte-Carlo method \cite{027} and many other 
techniques \cite{053,062}.

The coherent oscillations and current beats have been found in a short time scale response of 
a system upon abrupt change of the bias voltage.\cite{009,014} From the periods of the current 
beats it is possible to estimate the values of the QDs energy levels or the hopping parameters 
between them.\cite{038,051,058} The transient current characteristics can be also used to 
determine the spin relaxation time in some QD systems.\cite{004} Such phenomena have been 
investigated for QDs coupled to the normal leads as a result of the bias voltage pulse, \cite{005,022,029,031,036,041,053,061,076} driven by an
arbitrary time-dependent bias,\cite{026,027,040,053,062} by a sequence of rectangular pulses 
applied to the input lead \cite{017,032} or applied to the contact gradually switched on in 
time.\cite{025} The transient dynamics has been also studied for QD after a sudden symmetrical 
connection to the leads \cite{027,037,079,095} or asymmetrically coupled to electrodes following 
a sudden change of the QD energy levels.\cite{011} The transient heat generation driven by 
a step-like pulse bias with the Anderson-Holstein model or the time-dependent current through 
QD suddenly coupled to a vibrational mode have been studied in nanostructures with the normal 
\cite{019,030,047,056,064} or superconducting electrodes.\cite{072}

Technological progress in the real-time detection of single electrons has opened a possibility 
for studying electron transport from a perspective of the stochastic processes. Among theoretical 
tools for investigating the electron hopping statistics there are e.g. the full counting statistics 
(FCS) and the waiting time distribution (WTD)\cite{054,055,063,067,074,080}. These theoretical 
techniques  have been successfully applied to investigations of the transient processes via QD 
coupled to the normal leads\cite{063} or in hybrid systems with superconductors.\cite{067,074,080}  
Time-dependent processes are often investigated numerically, however, in exceptional cases some 
analytical results can give the deeper insight into considered problem. For instance, WTD in
the normal lead--QD--superconducting junction exhibit the coherent oscillations between the empty
and doubly occupied QD. \cite{074} Similarly,  some analytical calculations are possible for 
the energy transport in the polaronic regime described within the FCS method \cite{060}, for 
transient dynamics after a quench\cite{065}, for a phononic heat transport in the transient 
regime\cite{066} or for transient heat generation under a step-like bias pulse.\cite{044}

In this paper we analyze the sub-gap transport properties of a system comprising of a single 
QD which is tunnel coupled to: one metallic (normal) and two superconducting electrodes, 
focusing on transient effects driven by abrupt coupling of these constituents. It is natural 
that oscillations of the transient current would appear as a result of such quench, and the 
should depend on initial conditions of the system. Such  hybrid nanostructures with QD between 
the normal and superconducting electrodes, reveal many interesting effects with potential 
applications in nanoelectronics, spintronics or quantum computing. \cite{029,030,042,064,065} 
The superconducting reservoir affects the QD via proximity effect, and could be responsible 
for the Cooper-pair tunneling and Josephson currents, even in absence of any bias voltages. 
Additional normal electrode coupled to the system allows for good control of the electron 
transport \cite{Wernsdorfer_2012,Deblock_2015,Deblock_2016,Paaske_2015} 
and could significantly affect the transient phenomena.
Our goal is to investigate analytically the time-dependent QD occupation, the currents flowing from the normal and
superconducting leads, the induced QD pairing, the conductance and the time evolution of the Andreev bound states
(ABS).\cite{abs1,abs2,abs3,abs4,abs5,abs6} The formation of ABS signifies that superconducting correlations are
induced in the QD via the proximity effect. We investigate appearance in time of these states and study 
their spin-dependence. To perform analytical time-dependent calculations we assume that superconducting gap of 
both superconducting leads is the largest energy scale and we put it equal to infinity. Nevertheless, the realistic 
physics in the Andreev transport regime is still captured in this limit. Knowledge of the analytical formulas
allows us to find the answers to such questions as: (i) how do the considered quantities and their characteristics 
depend on the QD energy levels or the individual coupling of the QD with a given lead, (ii) what is the time period 
and frequency of these time-dependent quantities, and many related issues. Our investigations allow us 
also to analyze time evolution of the Andreev bound states and their dependence on the phase difference 
between the superconducting reservoirs. In our calculations we apply the equation of motion method for 
the second quantization operators and obtain their analytical form using the Laplace transform technique. 
Numerical calculations could provide results only for a specific choice of parameters and would not give 
deep insight into specific dependence of here considered quantities of our system. In this context 
the analytical calculations are much more general and could have some advantage over numerical data.

The paper is organized as follows. In Sec.~\ref{sec20} we present our model and discuss the theoretical 
formalism. The time-dependent QD occupancy is analyzed in Sec.~\ref{sec30}, whereas Sec.~\ref{sec40} is 
devoted to the proximity-induced pairing effects. The normal and superconducting transient currents 
through the QD are analyzed in Sec.~\ref{sec50} and in Sec.~\ref{sec60} we discuss the subgap 
conductance. In the last  Sec.~\ref{sec70} we draw the main conclusions of our study.

\section{\label{sec20}Model and theoretical description}

\begin{figure}
\centering
\includegraphics[width=60mm]{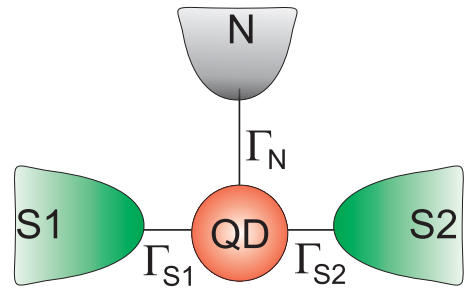}
\caption{\label{fig_schem} Schematic diagram for a quantum dot coupled with two superconducting leads (S1 and S2) and
one normal (metallic) electrode (N). }
\end{figure}

The system under consideration  consists of a QD placed between two superconducting leads ($S1$ and $S2$) and one
metallic electrode, $N$, see Fig.~\ref{fig_schem}.
The model Hamiltonian for this system can be written in the following form: $H=H_{S1}+H_{S2}+H_{N}+H_{QD}+H_{int}$,
where $H_{Sj}$ ($j=1,2$) describes electrons in the left or right superconducting lead
\begin{eqnarray}
H_{Sj} &=& \sum_{q\sigma} \varepsilon_{q_j,\sigma} c^+_{q_j \sigma} c_{q_j \sigma} + \sum_{q_j} \left( \Delta_j
c^+_{-q_j \uparrow } c^+_{q_j \downarrow } +h.c. \right) ,   \label{eq01}
\end{eqnarray}
$H_N$ refers to the normal lead, $H_N=\sum_{k\sigma} \varepsilon_{k\sigma} c^+_{k\sigma} c_{k\sigma}$, $H_{QD}$
describes the QD, $H_{QD}=\sum_{\sigma} \varepsilon_{\sigma} c^+_{\sigma} c_{\sigma}$. Electron transitions
between external leads and the QD are established by the tunnel Hamiltonian:
\begin{eqnarray}
H_{int}=\sum_{k, \sigma} V_{k\sigma} c_{k \sigma}^+c_{\sigma} + \sum_{j=1,2} \sum_{q \sigma} V_{q_j\sigma}
c^+_{q_j\sigma} c_{\sigma}+h.c.  \label{eq02}
\end{eqnarray}
We assume that the electron dispersion in all leads is spin-independent and impose the order parameters, $\Delta_j$, 
of the superconducting leads to be phase-dependent, $\Delta_j=|\Delta_j| \exp{(i\varphi_j)}$.  In our notation $k$
($q_j$) shall denote itinerant states of the normal (superconducting) lead. Correlations are neglected in our
calculations.

We are going to study time-response of this system on abrupt switching of the coupling parameters. We shall thus 
calculate the time-dependent QD occupations, $n_{\sigma}(t)$ and the currents flowing from the leads, $j_{N\sigma}(t)$,
$j_{S_j \sigma}(t)$. Additionally we will compute  $\langle c_{\downarrow}(t) c_{\uparrow}(t)\rangle$, which 
characterized the electron pairing induced at QD via proximity effect. In what follows we assume that all couplings
between the QD and the leads are suddenly switched on at $t=0^+$ (for $t\leq 0$ the QD is decoupled from the leads).
The time evolution of the considered quantities for $t>0$ depends on the initial QD filling and the chemical 
potentials. As time goes to infinity, we reproduce the stationary limit results known from the corresponding 
system. In this paper we use the Laplace transform method and our strategy in the
calculations is as follows: we construct the closed set of the equation of motion for creation and annihilation
operators (in the Heisenberg representation) $c_{\sigma}(t)$, $c_{k\sigma}(t)$, $c_{q_j\sigma}(t)$, $c^+_{\sigma}(t)$,
$c^+_{k\sigma}(t)$, $c^+_{q_j\sigma}(t)$,  using the Laplace transformations for these differential equations we obtain
the set of coupled algebraic forms  $c(s)=\int_0^{\infty} dt e^{-st}c(t)$ for all
considered operators. For instance, the QD occupation $n_{\sigma}(t)$ can be found from the relation
\begin{eqnarray}
n_{\sigma}(t)=  \langle \mathcal{L}^{-1}\{c^+_{\sigma}(s)\}(t) \cdot \mathcal{L}^{-1}\{c_{\sigma}(s)\}(t)
\rangle\label{eq03}
\end{eqnarray}
where $\mathcal{L}^{-1}\{a(s)\}(t)$ stands for the inverse Laplace transform of $a(s)$ and 
$\langle ... \rangle$ is the statistical averaging. 

Let us find the Laplace transforms of operators $c_{\sigma}(t)$ and $c_{q_j \sigma}(t)$ which are required 
to calculate the QD occupancy $\langle c_{\sigma}^{\dagger}(t) c_{\sigma}(t)\rangle \equiv n_{\sigma}(t)$, 
the QD induced pairing $\langle c_{\downarrow}(t) c_{\uparrow}(t)\rangle$ and the currents flowing from 
the leads. We write the  Laplace transformed equations of motions for the closed set of twelve operators 
(in the Heisenberg representation):
$c_{\uparrow}$, $c_{\downarrow}^{\dagger}$, $c_{k \uparrow}$, $c_{k \downarrow}^{\dagger}$, $c_{q_j \uparrow}$,
$c_{-q_j \downarrow}^{\dagger}$, $c_{q_j \downarrow}^{\dagger}$, $c_{-q_j \uparrow}$, $j=1,2$.
\begin{subequations} \label{EOM}
\begin{eqnarray} \label{EOM.A}
 (s+i\varepsilon_{\uparrow}) {c}_{\uparrow}(s) &=& -i \sum_{{r=k, q_1, q_2}} V_{r} {c}_{r\uparrow}(s)+{c}_{\uparrow}(0) , \\
\label{EOM.B}
 (s+i\varepsilon_{q_j}) {c}_{{q_j}\uparrow}(s) &=& -i V_{q_j} {c}_{\uparrow}(s) -i\Delta_j
{c}_{-{ q_j}\downarrow}^{\dagger}(s)+{c}_{{ q_j}\uparrow}(0) , \nonumber\\ \\
\label{EOM.C}
 (s-i\varepsilon_{q_j}) {c}_{-{ q_j}\downarrow}^{\dagger}(s) &=& i V_{q_j}
{c}_{\downarrow}^{\dagger}(s) -i\Delta_j^* {c}_{{q_j}\uparrow}(s)+{c}_{-{q_j}\downarrow}^{\dagger}(0) ,
\nonumber\\ \\
\label{EOM.D}
 (s+i\varepsilon_{k}) {c}_{{k}\uparrow}(s) &=& -i V_{k} {c}_{\uparrow}(s) +
{c}_{{k}\uparrow}(0) ,
\end{eqnarray}
\end{subequations}
\begin{subequations} \label{EOM2}
\begin{eqnarray} \label{EOM2.A}
 (s-i\varepsilon_{\downarrow}) {c}_{\downarrow}^{\dagger}(s) &=& i \sum_{{r=k, q_1, q_2}} V_{r} {c}_{r\downarrow}^{\dagger}(s)+{c}_{\downarrow}^{\dagger}(0) , \\
\label{EOM2.B}
 (s-i\varepsilon_{q_j}) {c}_{{q_j}\downarrow}^{\dagger}(s) &=& i V_{q_j} {c}_{\downarrow}^{\dagger}(s) -i\Delta_j^*
{c}_{-{q_j}\uparrow}(s)+{c}_{{q_j}\downarrow}^{\dagger}(0) , \nonumber\\ \\
\label{EOM2.C}
 (s+i\varepsilon_{q_j}) {c}_{-{ q_j}\uparrow}(s) &=& -i V_{q_j}
{c}_{\uparrow}(s) -i\Delta_j {c}_{{q_j}\downarrow}^{\dagger}(s)+{c}_{-{q_j}\uparrow}(0) ,
\nonumber\\ \\
\label{EOM2.D}
 (s-i\varepsilon_{k}) {c}_{{k}\downarrow}^{\dagger}(s) &=& i V_{k} {c}_{\downarrow}^{\dagger}(s) +
{c}_{{k}\downarrow}^{\dagger}(0) \,.
\end{eqnarray}
\end{subequations}
From Eqs.~\ref{EOM.A}-\ref{EOM.D} and Eqs.~\ref{EOM2.A}-\ref{EOM2.D} we get
\begin{subequations} \label{EOM3}
\begin{eqnarray} \label{EOM3.A}
{c}_{\uparrow}(s) M^{(+)}_{\uparrow}(s) &=&  {A}(s)-iK(s){c}_{\downarrow}^{\dagger}(s) \,, \\
 \label{EOM3.B}
 {c}_{\downarrow}^{\dagger}(s) M^{(-)}_{\downarrow}(s) &=&  {B}(s)-iK^*(s){c}_{\uparrow}(s) \,,
\end{eqnarray}
\end{subequations}
where
\begin{eqnarray}
K(s) &=& \sum_{j=1,2} { {V^2_{q_j} \Delta_j} \over {s^2+\varepsilon_{q_j}^2+|\Delta_j|^2}} \, ,
 \label{EOM3.C}
\end{eqnarray}
\begin{eqnarray}
A(s) &=& -\sum_{j=1,2} {{V_{q_j} } \left(\Delta_j c^+_{-q_j\downarrow}(0) +i (s - i\varepsilon_{q_j})c_{q_j\uparrow}(0)
\right) \over {s^2+\varepsilon_{q_j}^2+|\Delta_j|^2}}
 \nonumber\\
&&- i\sum_{k} {{V_{k} c_{k\uparrow}(0) } \over {s +i\varepsilon_{k}}}+c_{\uparrow}(0) \, ,
 \label{EOM3.D}
\end{eqnarray}
\begin{eqnarray}
{B}(s) &=& \sum_{j=1,2}  {{V_{q_j}} \left(\Delta_j^*
c_{-q_j\uparrow}(0)+i (s+i\varepsilon_{q_j})c^+_{qj\downarrow}(0)  \right) \over {s^2+\varepsilon_{q_j}^2+|\Delta_j|^2}}  \nonumber\\
&&+ i\sum_{k} {{V_{k} c^+_{k\downarrow}(0) } \over {s -i\varepsilon_{k}}}+c^+_{\downarrow}(0) \, ,
  \label{EOM3.E}
\end{eqnarray}
\begin{eqnarray}
M^{(+/-)}_{\sigma}(s) &=& s\pm i\varepsilon_{\sigma}+\sum_{j=1,2} {{V_{q_j}^2 (s \mp i \varepsilon_{q_j})}  \over
{s^2+\varepsilon_{q_j}^2+|\Delta_j|^2}} \nonumber\\ &+& \sum_{k} {{V_{k}^2 } \over {s \pm i\varepsilon_{k}}} \, .
 \label{EOM3.F}
\end{eqnarray}
Solving Eqs.~\ref{EOM3.A}, \ref{EOM3.B} we obtain for $c_{\uparrow}(s)$
\begin{eqnarray}
c_{\uparrow}(s)&=& \frac{M^{(-)}_{\downarrow}(s) A(s)-iK(s)B(s)}{M^{(+)}_{\uparrow}(s)
M^{(-)}_{\downarrow}(s)+K(s)K^*(s)} \, .
 \label{EOM4.A}
\end{eqnarray}
Repeating the same procedure to the set of operators: $c_{\downarrow}$, $c_{\uparrow}^{\dagger}$, $c_{k
\uparrow}^{\dagger}$, $c_{k \downarrow}$, $c_{q_j \downarrow}$, $c_{-q_j \uparrow}^{\dagger}$, $c_{q_j
\uparrow}^{\dagger}$ and $c_{-q_j \downarrow}^{\dagger}$ one can get
\begin{eqnarray}
c_{\downarrow}(s)&=&  \frac{M^{(-)}_{\uparrow}(s) B^+(s)+iK(s)A^+(s)}{M^{(-)}_{\uparrow}(s)
M^{(+)}_{\downarrow}(s)+K(s)K^*(s)} \,.  \label{EOM4.B}
\end{eqnarray}
Laplace transforms of $c_{\uparrow}^{\dagger}$ and $c_{\downarrow}^{\dagger}$ can be obtained, 
taking the hermitian conjugation of $c_{\uparrow}$ and $c_{\downarrow}$, respectively.

In the wide-band limit approximation and for $|\Delta_j| = \infty$  the functions $M^{+/-}_{\sigma}(s)$ and $K(s)$ can
be expressed in the following analytical forms:  $M^{+/-}_{\sigma}(s)=s\pm i \varepsilon_{\sigma} +\Gamma_N/2$, and
$K(s)=\left(\Gamma_{S_1}e^{i\varphi_1}+\Gamma_{S_2}e^{i\varphi_2} \right)/2$. Here we have assumed $\Gamma_{N/S_j}=2\pi
\sum_{k/q_j} V^2_{k/q_j} \delta(\varepsilon-\varepsilon_{k/q_j})$ and $\varepsilon_{k\sigma}=\varepsilon_k$,
$\varepsilon_{q_j\sigma}=\varepsilon_{q_j-\sigma}=\varepsilon_{-q_j}$. As an example, let us present explicit 
form of the Laplace transform for $c_{\uparrow}(t)$
\begin{eqnarray}
&& {c}_{\uparrow}(s) = \frac{1}{(s-s_{3})(s-s_{4})} \left\{ \left( s -
i\varepsilon_{\downarrow}+\frac{\Gamma_{N}}{2}\right)  \right.
 \label{EOM4.C} \\
 && \times \left[ {c}_{\uparrow}(0) -
i\sum_{k} \frac{V_{k} \; {c}_{{k}\uparrow}(0)} {s+i\varepsilon_{k}}  \right.
 \nonumber \\
 && \left.- \sum_{j=1,2}
\frac{i V_{q_j}(s-i\varepsilon_{q_j}){c}_{{q_j}\uparrow}(0)+V_{q_j}\Delta_j
{c}_{-{q_j}\downarrow}^{\dagger}(0)}{s^{2}+\varepsilon^{2}_{q_j}+|\Delta_j|^{2}} \right]
 \nonumber\\
 &-& \frac{i}{2} \left(\Gamma_{S1} e^{i\varphi_1}+ \Gamma_{S2}
e^{i\varphi_2}\right)
 \left[  {c}_{\downarrow}^{\dagger}(0) + i\sum_{k} \frac{V_{k} \; {c}_{{k}\downarrow}^{\dagger}(0)} {s-i\varepsilon_{ k}} \right. \nonumber
 \\
 && \left. \left. +  \sum_{j=1,2} \frac{i V_{q_j}(s+i\varepsilon_{q_j}){c}_{{q_j}
\downarrow}^{\dagger}(0) + V_{q_j}\Delta_j {c}_{-{q_j}\uparrow}(0)}{s^{2}+\varepsilon^{2}_{q_j}+|\Delta_j|^{2}} \right]
\right\} \,, \nonumber
\end{eqnarray}
where $s_{3,4}  =  \frac{1}{2} \left[- i(\varepsilon_{\uparrow}-\varepsilon_{\downarrow}) -\Gamma_{N}\pm i
\sqrt{\delta} \right]$,  $ \delta = (\varepsilon_{\uparrow}+\varepsilon_{\downarrow})^2 + \Gamma_{12}$ and
$\Gamma_{12}=\Gamma^2_{S_1}+\Gamma^2_{S_2}+2\Gamma_{S_1}\Gamma_{S_2} \cos(\varphi_1-\varphi_2)$.

Note, that in the formula (\ref{EOM4.C}) there appears the finite superconducting energy gap $\Delta_j$.
The limit $|\Delta_j|=\infty$ will be imposed later on, when computing the expectation values of
the product of two corresponding operators, e.g. $\langle c_{\sigma}^{\dagger}(t) c_{\sigma}(t)  \rangle$ or $\langle
c_{\sigma}^{\dagger}(t) c_{q_j\sigma}(t)  \rangle$. Additionally, expression for $c_{q_j \sigma}(s)$ needed 
for calculations of the currents flowing between the QD and the superconducting leads can be obtained from
Eqs.~\ref{EOM.B}, \ref{EOM.C}, \ref{EOM4.A}, \ref{EOM4.B} and it reads
\begin{eqnarray}
c_{q_j\sigma}(s) &=& \frac{1}{{s^2+\varepsilon_{q_j}^2+|\Delta_j|^2}} \left[  (s-i\varepsilon_{q_j}) (c_{q_j\sigma}(0)
-i V_{q_j} c_{\sigma}(s)) \right.
 \nonumber\\
 &+& \left. \alpha V_{q_j} \Delta_j c^+_{-\sigma}(s)-i\alpha \Delta_j c^+_{-q_j-\sigma}(0) \right] \,,
\label{EOM4.D}
\end{eqnarray}
where $\alpha=+(-)$ for $\sigma=\uparrow (\downarrow)$.  Using these formulas for $c_{\sigma}(s)$ and
$c_{q_j\sigma}(s)$ we can analytically determine the QD occupancy, pairing parameter, subgap currents 
and its differential conductance.

In the following we  set $e=\hbar=k_{B}\equiv 1$ and make use of the wide-band limit approximation. 
All numerical calculations shall be performed for $\Gamma_{S_1}=\Gamma_{S_2}=\Gamma_S$ and $\mu_N=0$, 
unless stated otherwise. The energies, currents and time are expressed in units of $\Gamma_{S}$, 
$e\Gamma_{S}/\hbar$ and $\hbar/\Gamma_{S}$, respectively. We assume the chemical potentials of 
superconducting leads $\mu_{S_1}=\mu_{S_2}=0$ to be grounded. For experimentally available values 
of $\Gamma_S$, $\Gamma_S \sim 200 \mu eV$ \cite{083,084,085}
the typical time and current units would be $\sim 3.3 psec$ and $\sim 48 nA$, respectively.

\section{\label{sec30}Quantum dot occupancy}

\begin{figure}[b!]
\centering
\includegraphics[width=80mm]{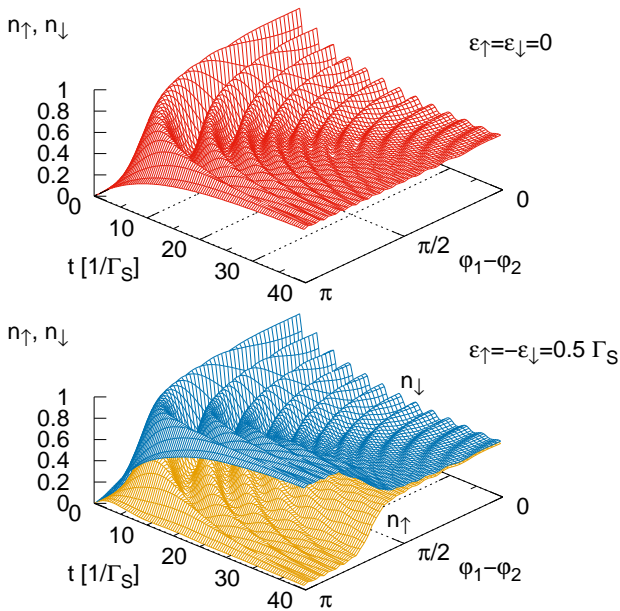}
\caption{\label{fig01} Time evolution of the QD occupancies $n_{\uparrow}(t)$, $n_{\downarrow}(t)$ as a function of the
phase difference $\phi$ for $\varepsilon_{\sigma}=0$ (upper panel) and
$\varepsilon_{\uparrow}=-\varepsilon_{\downarrow}=0.5$ (bottom panel). $\Gamma_{S1}=\Gamma_{S2}=\Gamma_S=1$,
$\Gamma_N=0.1$, $\mu_N=0$, $|\Delta_1|=|\Delta_2|\rightarrow \infty$, $n_{\sigma}(0)=0$. The QD occupancies satisfy the
relation $n_{\sigma}(\phi)=n_{\sigma}(\phi+2\pi)$ and for $t=\infty$ are symmetrical  regard to $\phi=\pi$.
}
\end{figure}

Let us consider the time-dependent QD occupancy after abrupt  coupling (at $t=0^+$) to 
the normal and superconducting electrodes. We assume no bias voltage between electrodes 
and make use of the wide band limit approximation and impose $|\Delta_j|=\infty$. Under 
these assumptions the QD occupation, $n_{\sigma}(t)$, reads  (cf. \cite{081} for N-QD-S 
and \cite{090} for N-QD-N systems):
\begin{widetext}
\begin{eqnarray}\label{eq06}
n_{\sigma}(t)&=&   \mathcal{L}^{-1}\left\{ {s+i\varepsilon_{-\sigma}+\Gamma_N/2 \over (s-s_1)(s-s_2)} \right\}(t)
\mathcal{L}^{-1}\left\{ {s-i\varepsilon_{-\sigma}+\Gamma_N/2 \over (s-s_3)(s-s_4)} \right\}(t) n_{\sigma}(0)
  \\
  &+& {\Gamma_{12}\over 4 }\mathcal{L}^{-1}\left\{ {1 \over (s-s_1)(s-s_2)} \right\}(t)
\mathcal{L}^{-1}\left\{ {1 \over (s-s_3)(s-s_4)} \right\}(t) (1-n_{-\sigma}(0)) \nonumber\\
&+& \sum_{k_1,k_2} V_{k_1} V_{k_2}  \mathcal{L}^{-1}\left\{ {s+i\varepsilon_{-\sigma}+\Gamma_N/2 \over
(s-s_1)(s-s_2)(s-i\varepsilon_{k_1})} \right\}(t)
\mathcal{L}^{-1}\left\{ {s-i\varepsilon_{-\sigma}+\Gamma_N/2 \over (s-s_3)(s-s_4)(s+i\varepsilon_{k_2})} \right\}(t)
\left\langle   c^+_{k_1 \sigma}(0) c_{k_2 \sigma}(0)
 \right\rangle  \nonumber\\
 &+& {\Gamma_{12}\over 4 } \sum_{k_1,k_2} V_{k_1} V_{k_2}   \mathcal{L}^{-1}\left\{ {1 \over
(s-s_1)(s-s_2)(s+i\varepsilon_{k_1})} \right\}(t)
   \mathcal{L}^{-1}\left\{ {1 \over
(s-s_3)(s-s_4)(s-i\varepsilon_{k_2})}  \right\}(t) \left\langle c_{k_1-\sigma}(0)c^+_{k_2-\sigma}(0)  \right\rangle \,
, \nonumber
\end{eqnarray}
where $s_{1,2}={1 \over 2}\left[i(\varepsilon_{\uparrow}-\varepsilon_{\downarrow})-\Gamma_N \pm i\sqrt{\delta}
\right]$, and for $\sigma=\downarrow$ one should replace $(s_1,s_2) \leftrightarrow (s_3,s_4)$, respectively. The first
two terms describe the transient QD charge oscillations which  depend on the initial QD occupations. The last two terms
(with the sums over $k$) are related to the normal lead and they give non-vanishing and non-oscillating contribution to
$n_\sigma(t)$, regardless of the initial conditions. Note that in Eq.~\ref{eq06}  the  terms involving the
expectation values of the product of electron annihilation and creation operators $c_{q_j \sigma}$ and 
$c_{q_j \sigma}^{\dagger}$ of the superconducting lead electrons do not appear. Such terms take e.g. 
the following integral form  (cf. \cite{081}):
\begin{eqnarray}
&& \frac{\Gamma_S}{2\pi} \int_{-\infty}^{+\infty} d\varepsilon f_S(\varepsilon) {\cal{L}}^{-1} \left\{ \frac{\left(
s+i\varepsilon_{\downarrow}+\frac{\Gamma_{N}}{2}\right)\left(s+i\varepsilon \right)}{(s-s_1)(s-s_2)(s^2+\varepsilon^2 +
|\Delta_j|^2) } \right\}(t)   {\cal{L}}^{-1} \left\{ \frac{\left(
s-i\varepsilon_{\downarrow}+\frac{\Gamma_{N}}{2}\right)\left(s-i\varepsilon \right)}{(s-s_3)(s-s_4)(s^2+\varepsilon^2 +
|\Delta_j|^2) } \right\}(t) \,,
\end{eqnarray}
where $f_s(\varepsilon)$  is the Fermi distribution function. It is easy to check
numerically that the above integral over the energy is smaller and smaller with increasing $|\Delta_j|$. Thus in our
calculations for $|\Delta_j| = \infty$ we can neglect all terms involving operators $\hat{c}_{{\bf q} \sigma}(0)$. The
formula~\ref{eq06} can be further elaborated and after some algebra one rewrites the two first
terms explicitly while the third and fourth terms can be expressed by integrals over the energy in the
normal lead spectrum
\begin{eqnarray}\label{eq08}
n_{\sigma}(t)&=& e^{-\Gamma_N t}  \left[n_{\sigma}(0)+ \left(1-n_{\sigma}(0)-n_{-\sigma}(0)\right)
\sin^2\left(\frac{\sqrt{\delta}t}{2}\right) \frac{\Gamma_{12}}{\delta} \right]
 \\
&+& {\Gamma_N \over 2 \pi} \int_{-\infty}^{+\infty} d\varepsilon f_N(\varepsilon) \mathcal{L}^{-1}\left\{
{s+i\varepsilon_{-\sigma}+\Gamma_N/2 \over (s-s_1)(s-s_2)(s-i\varepsilon)} \right\}(t)
 \cdot \mathcal{L}^{-1}\left\{ {s-i\varepsilon_{-\sigma}+\Gamma_N/2 \over (s-s_3)(s-s_4)(s+i\varepsilon)} \right\}(t)
  \nonumber\\
&+& {\Gamma_N  \over 8 \pi} \Gamma_{12}  \int_{-\infty}^{+\infty} d\varepsilon \left(1- f_N(\varepsilon)\right)
\mathcal{L}^{-1}\left\{ {1 \over (s-s_1)(s-s_2)(s+i\varepsilon)  }\right\}(t)  \cdot \mathcal{L}^{-1}\left\{ {1 \over
(s-s_3)(s-s_4)(s-i\varepsilon) }\right\}(t) . \nonumber
\end{eqnarray}
\end{widetext}
Here $f_N(\varepsilon)$ is the electron Fermi distribution function for the normal lead and for $\sigma ={\downarrow}$
the replacement $(s_1,s_2) \leftrightarrow (s_3,s_4)$ should be done.  The phase difference $\phi$ enters 
Eq.~\ref{eq08} only through the function $\cos{\phi}$, therefore the QD occupancy satisfies the symmetry relation
$n_{\sigma}(\phi)=n_{\sigma}(\phi+2\pi)$. Note that the part
which depends on the initial QD filling oscillates with the period ${2\pi / \sqrt{\delta}}$. These oscillations depend
on the QD electron energies, $\varepsilon_{\uparrow}+\varepsilon_{\downarrow}$, both couplings $\Gamma_{S_1}$,
$\Gamma_{S_2}$ and the phase difference $\phi$ of the superconducting order parameters, $\phi=\varphi_1-\varphi_2$. The
oscillations are damped due to the exponential factor $e^{-\Gamma_N t}$ and in the asymptotic time limit the
information about the initial QD occupation is entirely washed out. From Eq.~\ref{eq08} we infer that, when 
QD is coupled only to the superconducting leads and the initial conditions are $n_{\sigma}(0)=(1,0)$ or $(0,1)$, 
the time-dependent QD occupancy does not change at all (independently of $\phi$ and $\Gamma_{S_{1/2}}$). 
In this case the QD is occupied only be one electron which cannot be exchanged with the superconducting 
reservoirs due to the infinity large energy gaps. For the initial conditions $n_{\sigma}(0)=(1,1)$ or $(0,0)$ 
the oscillations of the QD occupancy oscillates with the time period $T=\frac{2\pi}{\sqrt{\delta}}$ for 
$\phi \neq \pi$ independently of $\Gamma_{S_{1/2}}$ or for $\phi=\pi$, $\Gamma_{S_{1}} \neq \Gamma_{S_2}$. 
These oscillations, however, disappear for $\phi=\pi$ and $\Gamma_{S_{1}} = \Gamma_{S_2}$ as shown 
in Fig.~\ref{fig01}.

%
%

The formula (\ref{eq08}) for $\Gamma_N=0$ resembles the Rabbi oscillations of a typical two-level quantum system
described by the effective Hamiltonian $H_{eff}=\frac{1}{2} \left(\Gamma_{S_1}e^{i \varphi_1}+\Gamma_{S_2}e^{i
\varphi_2} \right) c_{\uparrow}^{\dagger}  c_{\downarrow}^{\dagger} + h.c. + \sum_{\sigma} \varepsilon_{\sigma}
n_{\sigma}$. Assuming that at $t=0$ the QD is empty, $n_{\sigma}(0)=0$, we can calculate the probability $P(t)$ of
finding the QD in the doubly occupied configuration , $n_{\uparrow}=n_{\downarrow}=1$. Within the standard treatment of
a two-level system we have\cite{081,Rabbi}
\begin{eqnarray}\label{Rabbi}
P(t)=\frac{\Gamma_{12}}{\Gamma_{12}+(E_1-E_2)^2} \sin^2\left(\sqrt{\Gamma_{12}+(E_1-E_2)^2} \frac{t}{2}  \right) ,
\nonumber\\
\end{eqnarray}
where $E_1=0$ and $E_2=\varepsilon_{\uparrow}+\varepsilon_{\downarrow}$ are energies of the empty and double 
occupied configurations, respectively. This formula can be rewritten as $P(t)=\frac{\Gamma_{12}}{\delta} 
\sin^2\left( \frac{\sqrt{\delta}}{2} t \right)$ and becomes identical with our expression (\ref{eq08}) 
obtained for $n_{\sigma}(0)=0$, $\Gamma_N=0$.

To illustrate such analytical results and to reveal influence of the phase difference of two superconducting 
leads on the QD occupation in Fig.~\ref{fig01} we show $n_{\uparrow}(t)$ and  $n_{\downarrow}(t)$ with respect 
to time and $\phi$ for  $\varepsilon_{\sigma}=0$ (upper panel) and for the Zeeman splitting
$\varepsilon_{\sigma}=-\varepsilon_{-\sigma}=0.5$ (bottom panel). We consider here the symmetric coupling
$\Gamma_{S_1}=\Gamma_{S_1}=\Gamma_{S}$ and assume the initial conditions $n_{\sigma}(0)=(0,0)$. Note that 
for $\varepsilon_{\sigma}=0$ the QD occupancy becomes spin-independent, i.e. $n_{\sigma}(t)=n_{-\sigma}(t)$  
(see Eq.~\ref{eq08}). For $t \rightarrow \infty$ it always tends to $0.5$, regardless of the superconducting 
phase difference. In absence of any phase--difference we observe the oscillations of $n_{\sigma}(t)$ with 
the period $T={\pi / \Gamma_S}$ which are damped according to the exponential function $e^{-\Gamma_N t}$. 
Notice, that period of these oscillations is twice shorter compared to the oscillations in the N-QD-S 
system.\cite{081} For $\phi\neq 0$ these oscillations are characterized by the phase-dependent period 
$T={\pi / [\Gamma_S |\cos(\phi/2)|]}$. For the special case $\phi=\pi$ ($\Gamma_{12}=0$) the oscillations 
disappear and the QD charge develops in time exactly in the same way as for the QD coupled only to 
the normal lead (with $\varepsilon_{\sigma}=0$), e.g. \cite{001}:
\begin{eqnarray}\label{eq08cc}
n_{\sigma}(t)&=& n_{\sigma}(0) e^{-\Gamma_N t} \\ &+& {\Gamma_N \over \pi} e^{-\Gamma_N t/2}
\int_{-\infty}^{+\infty}d\varepsilon  f_N(\varepsilon)  {\cosh(\Gamma_Nt/2)-\cos(\varepsilon t) \over
(\Gamma_N/2)^2+\varepsilon^2}  \,. \nonumber
\end{eqnarray}
For $\mu=0$ and the zero temperature case we obtain $n_{\sigma}(t)={1\over 2} + e^{-\Gamma_N t}
\left(n_{\sigma}(0)-{1\over 2} \right)$. It means that for $n_{\sigma}(0)=(0,0)$ or $(1,1)$ 
the QD occupation increases or decreases monotonically in time without any oscillations, 
changing from zero (one) to 0.5 (see Fig.~\ref{fig01}, upper panel).

The situation changes in the presence of the Zeeman splitting (bottom panel). For symmetric 
splitting of around $\mu_N=0$, $\varepsilon_{\uparrow}=-\varepsilon_{\downarrow}$, the first 
term of Eq.~\ref{eq08} depends only on the phase difference $\phi$ and $\Gamma_{S_i}$. Its 
contribution to the final QD occupancy is the same for arbitrary values of $\varepsilon_{\sigma}$. 
On the other hand the two last  terms in Eq.~\ref{eq08} depend separately on 
$\varepsilon_{\sigma}$. For $\phi=\pi$ the contribution from these terms 
is identical with the case of the QD  coupled only to the normal lead. For $t=40$ and
$\varepsilon_{\uparrow}=-\varepsilon_{\downarrow}=0.5$ (bottom panel in Fig.~\ref{fig01}), 
the contribution from $\uparrow$ ($\downarrow$) is $\sim 0.03$ ($\sim 0.95$). 
For $\phi=0$ such contributions become $\sim 0.49$ and $\sim 0.51$, respectively. 
One can thus control the QD occupancy by means of the phase difference parameter $\phi$.

%
\begin{figure}
\centering
\includegraphics[width=65mm]{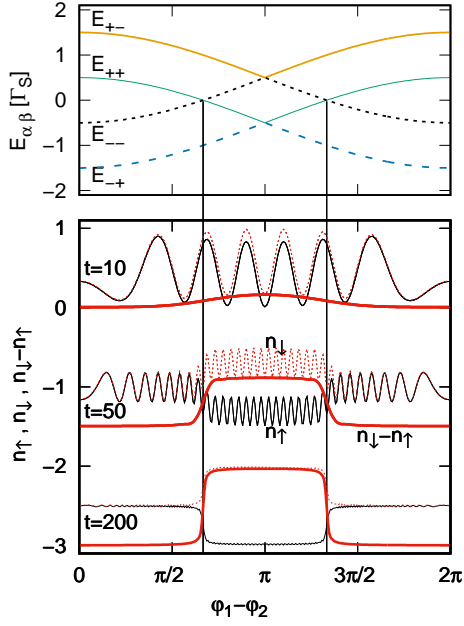}
\caption{\label{fig03new} Energies of the Andreev bound states of the proximitized QD, $E_{\alpha \beta}$, (upper
panel) and QD occupancies:  $n_{\uparrow}$, $n_{\downarrow}$ and $n_{\downarrow}-n_{\uparrow}$ (solid black, broken red
and thick red curves, respectively) as a function of $\phi$. The occupancies are obtained for $t=10$, $t=50$ (shifted
down by 1.5) and for $t=200$ u.t. (shifted down by 3.0). The vertical black lines indicate characteristic points for
$\phi=\frac{2\pi}{3}$ and $\phi=\frac{4\pi}{3}$, and the other parameters are
$\varepsilon_{\uparrow}=-\varepsilon_{\downarrow}=0.5$, $\Gamma_{S1}=\Gamma_{S2}=\Gamma_S=1$, $\Gamma_N=0.02$,
$\mu_N=0$. }
\end{figure}
%
Let us analyze more carefully variation of the QD occupancy against the phase difference $\phi$. 
In Fig.~\ref{fig03new} (upper panel) we present the ABS energies of the proximitized QD, $E_{\alpha
\beta}=\bar{E}_{\alpha}-\varepsilon_{\beta}$, ($\alpha= \pm$, $\beta = \pm \equiv \uparrow/\downarrow$),
where $\bar{E}_{\alpha}=  \frac{1}{2} (\varepsilon_{\uparrow}+\varepsilon_{\downarrow}) + \alpha
\sqrt{\frac{(\varepsilon_{\uparrow}+\varepsilon_{\downarrow})^2}{4} +\Gamma_S^2 \cos^2\frac{\phi}{2} }$ 
is the quasiparticle energy representing a superposition of the empty and double occupied states~\cite{096}. 
In the lower panel we show the QD occupancies $n_{\uparrow}(t)$, $n_{\downarrow}(t)$ and the difference
$n_{\downarrow}(t)-n_{\uparrow}(t)$ for $\Gamma_N=0.02$ obtained for particular times $t$. 
QD occupancy rapidly changes for such values of $\phi$ which satisfy the relation $E_{++}=E_{--}$, 
i.e. for $\phi=\pi\pm \arccos{\frac{\varepsilon_{\uparrow}}{\Gamma_S}}$ (here 
$\varepsilon_{\uparrow}+\varepsilon_{\downarrow}=0$, $\varepsilon_{\uparrow}>0$).  
Exactly for such values of $\phi$ we observe the transition $0-\pi$ in our system. 
This transition is clearly visible in the long time (steady) limit. 
In our case for $\Gamma_N=0.02$ this time equals $~200$ u.t. (approximately equal to
$\frac{4}{\Gamma_N}$). For greater $\Gamma_N$, $\Gamma_N=0.1$, such transition is observed 
(although it is smeared around $\phi=\pi \pm \pi/3$) already for $t~40$ u.t., see 
the lower panel in Fig.~\ref{fig01}. At very early stage of the time evolution 
such $0-\pi$ transition is only weakly manifested by the time-dependent magnetization
$n_{\downarrow}(t)-n_{\uparrow}(t)$. On the other hand, oscillations of the QD occupancies 
hardly detect existence of this transition. However, already for $t\simeq\frac{1}{\Gamma_N}=50$ 
u.t. this transition is well marked on the occupancy curves as well as on $n_{\downarrow}(t)
-n_{\uparrow}(t)$. Notice the decreasing amplitude and increasing frequency of the QD 
occupancies versus time. These transient characteristics are described by the factor 
$\sin^2(2\Gamma_S |\cos(\phi/2)|t)e^{-\Gamma_N t}$, see the first term of Eq.~\ref{eq08}. 
Let us emphasize, that despite oscillatory character of $n_{\sigma}(t)$, the resulting 
magnetization $n_{\downarrow}(t)-n_{\uparrow}(t)$ is a smooth function of $\phi$.

\section{\label{sec40}Induced on-dot pairing}

We shall now calculate the pairing amplitude  $\chi(t) \equiv \langle c_{\downarrow}(t)
c_{\uparrow}(t)\rangle$ driven by the proximity effect, assuming absence of any bias 
voltage ($\mu_N=0$).  Using the expressions for $c_{\uparrow}(s)$ and $c_{\downarrow}(s)$
obtained in Sec.~\ref{sec20} we find
%
%

\begin{eqnarray}\label{eq13}
\chi(t) &=&  -{i \over 2}\left(\Gamma_{S1}e^{i\varphi_1}+\Gamma_{S2}e^{i\varphi_2}  \right) \times
 \\
 &&\left[ - n_{\uparrow}(0) \mathcal{L}^{-1}\left\{ {1 \over (s-s_1)(s-s_2)} \right\}(t) \right.
 \nonumber\\ &&
\mathcal{L}^{-1}\left\{ {s-i\varepsilon_{\downarrow}+\Gamma_N/2 \over (s-s_3)(s-s_4)} \right\}(t)
 \nonumber\\
  &+&   (1-n_{\downarrow}(0))
\mathcal{L}^{-1}\left\{ {s-i\varepsilon_{\uparrow}+\Gamma_N/2 \over (s-s_1)(s-s_2)} \right\}(t)
 \nonumber\\ &&
 \left. \mathcal{L}^{-1}\left\{{1 \over (s-s_3)(s-s_4)} \right\}(t) + \frac{\Gamma_N}{2 \pi} \Phi_{\uparrow}^*
  \right]  \nonumber
\end{eqnarray}
%
%
where
\begin{eqnarray}\label{eq13ee}
\Phi_{\sigma} &=&  \int_{-\infty}^{+\infty} d\varepsilon  \mathcal{L}^{-1}\left\{ {1 \over
(s-s_1)(s-s_2)(s+i\varepsilon)  }\right\}(t)
 \\ &&
  \mathcal{L}^{-1}\left\{
{s+i\varepsilon_{\sigma}+\Gamma_N/2 \over (s-s_3)(s-s_4)(s-i\varepsilon)  }\right\}(t) (1-f_N(\varepsilon))
 \nonumber\\ &-&
  \int_{-\infty}^{+\infty} d\varepsilon f_N(\varepsilon) \mathcal{L}^{-1}\left\{
{s+i\varepsilon_{-\sigma}+\Gamma_N/2 \over (s-s_1)(s-s_2)(s-i\varepsilon) }\right\}(t)
 \nonumber\\ &&
\mathcal{L}^{-1}\left\{ {1 \over (s-s_3)(s-s_4)(s+i\varepsilon)  }\right\}(t) ,   \nonumber
\end{eqnarray}
and the replacement $(s_1,s_2) \rightarrow (s_3,s_4)$ should be made for $\sigma=\downarrow$. 
The terms proportional to $n_{\uparrow}(0)$ and $(1-n_{\downarrow}(0))$ in the above relation 
can be expressed analytically, and 
\begin{eqnarray}\label{eq13b}
\chi(t) &=& {i \over 2} \left(\Gamma_{S_1}e^{i\varphi_1}+\Gamma_{S_2}e^{i\varphi_2}  \right)
 \\
 && \left\{ -\frac{\Gamma_N}{2 \pi} \Phi_{\uparrow}^* + \left(n_{\downarrow}(0)+n_{\uparrow}(0)-1\right) e^{-\Gamma_N
 t} \right.
  \nonumber\\
&& \left. \times \left[ \sqrt{\delta} \sin\left(\sqrt{\delta} t\right) +  i
(\varepsilon_{\uparrow}+\varepsilon_{\downarrow}) (\cos(\sqrt{\delta} t)-1)\right]/\delta \right\} \nonumber  \,.
\end{eqnarray}
%
\begin{figure}
\centering
\includegraphics[width=70mm]{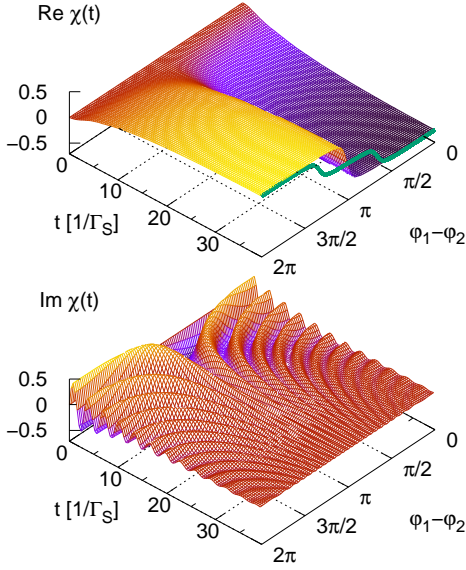}
\caption{\label{fig03} The real (upper panel) and imaginary  (bottom panel) parts of the QD induced pairing   $\chi(t)
= \langle c_{\downarrow}(t)c_{\uparrow}(t)\rangle$ as a function of time and the phase difference $\varphi_1-\varphi_2$
for $\varepsilon_{\sigma}=0$, $\Gamma_{S_1}=\Gamma_{S_2}=\Gamma_S$, $\Gamma_N=0.1$ and $n_{\sigma}(0)=0$.  $\chi(t)$
satisfies the relation $\chi(t,\phi)=\chi(t,\phi+4\pi)$ and for $t=\infty$ is symmetrical about $\phi=2\pi$. The bold
green line for $t=40$ in the upper panel corresponds to the case
$\varepsilon_{\uparrow}=-\varepsilon_{\downarrow}=0.5$.
}
\end{figure}
Notice, that for $\Gamma_{S_1}=\Gamma_{S_2}=\Gamma_S$ and $\phi =\pi$  the factor
$\Gamma_{S_1}e^{i\varphi_1}+\Gamma_{S_2}e^{i\varphi_2} =2\Gamma_S \cos{\phi \over 2}$ 
vanishes therefore the on-dot pairing $\langle c_{\downarrow}(t)c_{\uparrow}(t) \rangle$ 
is absent (see upper and bottom panels in Fig.~\ref{fig03} for $\phi=\pi$). However, 
for  $\phi \neq \pi$ and $\varepsilon_{\uparrow}
+\varepsilon_{\downarrow}=0$ we have
\begin{eqnarray}\label{eq13b}
\chi(t) &=& -{\Gamma_N \Gamma_S \over 2 \pi} \cos{\frac{\phi}{2}} \Im \Phi_{\uparrow} +
\frac{i}{2} \left(n_{\downarrow}(0)+n_{\uparrow}(0)-1\right)  \nonumber \\ &\times & e^{-\Gamma_N t}
\frac{\cos{\frac{\phi}{2}}}{|\cos{\frac{\phi}{2}}|}
 \sin\left(2 \Gamma_S |\cos{\frac{\phi}{2}}| t \right) \,,
\end{eqnarray}
where we have used the obvious property $\Re \Phi_{\sigma} =0$, corresponding to $\mu_N=0$ \cite{081}. 
The imaginary part of $\chi(t)$ oscillates with the same period as the QD occupancy, 
i.e. with  $T=\pi /[{\Gamma_S |\cos(\phi/2)}|]$, but the real part changes monotonically 
from zero to some constant  value without any oscillations (upper panel). We can notice, that
the imaginary part of $\chi$ vanishes when the QD is filled by a single electron at the initial 
time $t=0$. On the other hand, the real part of $\chi$ is a non-vanishing function irrespective 
of the initial conditions. It is worth mentioning, that for $\Gamma_N=0$ (i.e. Josephson 
junction setup) and for $\varepsilon_{\uparrow}+\varepsilon_{\downarrow}=0$  the expression 
for $\chi(t)$ becomes purely imaginary and is characterized by undamped oscillations inducing 
d.c. current (see the next section \ref{sec50}). In general, from the analysis of Eq.~\ref{eq13} 
we infer that the QD induced pairing satisfies the symmetry relation $\chi(t,\phi)
=\chi(t,\phi+4\pi)$. In particular, for $t=\infty$, it becomes symmetric with respect 
to $\phi=2\pi$.

\section{\label{sec50}Subgap currents}

Let us consider the currents $j_{N\sigma}(t)$ and $j_{Sj\sigma}(t)$ flowing between the QD 
and the normal or superconducting electrodes, respectively. These currents depend on time 
due to the abrupt coupling of all parts of the considered system. For $t>0$, even at zero 
bias voltage,  there are induced transient currents. Such electron currents can be obtained 
from the evolution of the total number of electrons of the corresponding electrode \cite{001}. 
For the normal lead we can express it as\cite{038,050,051,058,069}:
\begin{eqnarray}\label{eq09}
j_{N\sigma}(t)&=& 2 \Im \left( \sum_k V_{k\sigma} e^{-i \varepsilon_{k\sigma}t} 
\langle c^+_{\sigma}(t)c_{k\sigma}(0) \rangle \right) \nonumber\\ 
&-& \Gamma_N n_{\sigma}(t) \,,
\end{eqnarray}
where we have assumed the energy-independent  normal lead spectrum. Using the formulas 
of Sec.~\ref{sec20} we find 
\begin{eqnarray}\label{eq10}
j_{N \sigma}(t)&=& {\Gamma_N \over \pi} \Re \left(  \int_{-\infty}^{+\infty} d\varepsilon f_N(\varepsilon)
e^{-i\varepsilon t} \right.
 \\ &&
 \left. \mathcal{L}^{-1}\left\{ {s+i\varepsilon_{-\sigma}+\Gamma_N/2 \over
(s-s_1)(s-s_2)(s-i\varepsilon) }\right\}(t) \right) \nonumber\\ &&-\Gamma_N n_{\sigma}(t) \nonumber .
\end{eqnarray}
%
%
\begin{figure}[b!]
\centering
\includegraphics[width=70mm]{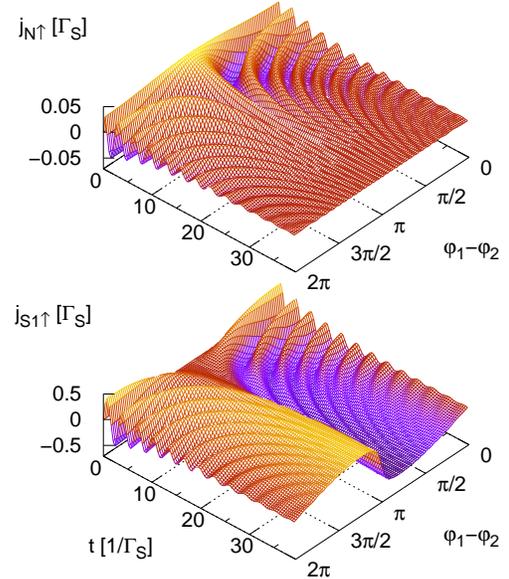}
\caption{\label{fig02} The time dependent currents flowing between the QD and the normal lead, $j_{N\uparrow}(t)$,
(upper panel) and between the QD and the superconducting lead, $j_{S1\uparrow}(t)$, (bottom panel), as a function of
the phase difference $\varphi_1-\varphi_2$. The system parameters are: $\varepsilon_{\sigma}=0$,
$\Gamma_{S_1}=\Gamma_{S_2}=\Gamma_S$, $\Gamma_N=0.1$ and $n_{\sigma}(0)=0$. }
\end{figure}
%
Inserting the inverse Laplace transform  and using the expression for $n_{\sigma}(t)$ one can obtain
analytical relation for $j_{N \sigma}(t)$. However, this solution for arbitrary $t$ cannot be written 
in relatively compact (or transparent) form, so we restrict ourselves to the asymptotics
$t=\infty$
\begin{eqnarray}\label{eq10bb}
j_{N \sigma}&=& {\Gamma_N \over \pi}   \int d\varepsilon \left\{  f_N(\varepsilon) \left[ \Re \left(
\frac{i(\varepsilon+\varepsilon_{-\sigma})+\frac{\Gamma_N}{2}} { \left({\Gamma_N \over 2} +i \varepsilon_{++}\right)
\left({\Gamma_N \over 2} +i \varepsilon_{-+}\right) } \right)  \right. \right.
 \nonumber\\
 && \left. \,\,\, - {\Gamma_N \over 2}  \frac{(\varepsilon+\varepsilon_{-\sigma})^2+\frac{\Gamma_N^2}{4}} { \left({\Gamma_N^2 \over 4} +
\varepsilon_{++}^2\right)\left({\Gamma_N^2 \over 4} + \varepsilon_{-+}^2\right) } \right]
 \\
 &-& \left.  (1-f_N(\varepsilon)) \frac {\Gamma_N \Gamma_{12}} {8 \left({\Gamma_N^2 \over 4} +
\varepsilon_{+-}^2\right) \left({\Gamma_N^2 \over 4} + \varepsilon_{--}^2\right) }  \right\} \nonumber ,
\end{eqnarray}
where $\varepsilon_{\alpha \beta}=\varepsilon + E_{\alpha \beta}$ and $E_{\alpha \beta}$ are 
the quasiparticle energies of the proximitized QD.

Fig.~\ref{fig02} (upper panel) shows  the time-dependent current flowing from the normal lead
to the QD as function of the phase difference $\phi=\varphi_1-\varphi_2$ obtained for the unbiased 
system $\varepsilon_{\sigma}=0$. At a beginning the current starts to flow from the normal electrode
to the empty QD. In a next stage, electrons tunnel in both directions with the characteristic 
oscillations. These damped oscillations are clearly visible and for $t \rightarrow \infty$ 
the current vanishes for all $\phi$. The period of these oscillations increases with $\phi$, 
similarly to the behavior observed for the QD occupancy. Exceptionally, for $\phi=\pi$, 
the current tends to its asymptotic value without any oscillations according to the formula 
(valid for the zero temperature, $\varepsilon_{\sigma}=0$ and $\Gamma_{S_1}=\Gamma_{S_2}$):
\begin{eqnarray}\label{eq12}
j_{N \sigma}(t)&=& {\Gamma_N}  e^{-\Gamma_N  t} \left( {1 \over 2}-n_{\sigma}(0) \right) \,.
\end{eqnarray}
We can notice, that right after the abrupt coupling (at $t=0^+$) the large value of 
transient current $j_{N\sigma}$ is induced in the system ($\sim \frac{\Gamma_N}{2}$) 
which is artifact of the WBL approximation. \cite{015} We have checked that by applying 
a more realistic (smooth) QD-leads coupling profile the initial current would gradually
increase, revealing the same period of oscillations and other overall features.\cite{081}

The situation looks a bit different for the currents, flowing between the QD and 
superconducting leads. To calculate these currents we start from the standard 
formula $j_{Sj\sigma}(t)= 2 \Im \left( \sum_{qj} V_{qj} \langle
c^+_{\sigma}(t)c_{qj\sigma}(t) \rangle \right)$ and use the Laplace transforms 
for $c^+_{\sigma}(s)$ and $c_{qj\sigma}(s)$, obtaining\cite{081}
\begin{widetext}
\begin{eqnarray}\label{eq13cc2}
j_{S_{1/2} \sigma}(t) &=& \Re \left\{ {\Gamma_{S_{1/2}} \over 2} \left(\Gamma_{S_{1/2}}+\Gamma_{S_{2/1}}e^{\pm i \phi} \right) \right. \times \\
 && \left[ \frac{\Gamma_N}{2 \pi} \Phi_{\sigma} - n_{\sigma}(0) \mathcal{L}^{-1}\left\{ {1 \over (s-s_3)(s-s_4)} \right\}(t)
\mathcal{L}^{-1}\left\{ {s+i\varepsilon_{-\sigma}+\Gamma_N/2 \over (s-s_1)(s-s_2)} \right\}(t)  \right.
 \nonumber\\ &+& \left. \left. (1-n_{-\sigma}(0)) \mathcal{L}^{-1}\left\{{1 \over (s-s_1)(s-s_2)} \right\}(t)
\mathcal{L}^{-1}\left\{ {s+i\varepsilon_{\sigma}+\Gamma_N/2 \over (s-s_3)(s-s_4)} \right\}(t) \right] \right\}
  \,, \nonumber
\end{eqnarray}
\end{widetext}
As usually, the replacement $(s_1,s_2) \leftrightarrow (s_3,s_4)$ should be made for 
$\sigma=\downarrow$.  After straightforward algebra we can derive more explicit form 
for the superconducting current
\begin{eqnarray}\label{eq13dd2}
j_{S_{1/2} \sigma}(t)&=& {\Gamma_{S_{1/2}} \over 2 \delta} \left(1-n_{\sigma}(0)-n_{-\sigma}(0)\right) e^{-\Gamma_N t}
 \\
&& \left[ (\Gamma_{S_{2/1}}\cos\phi +\Gamma_{S_{1/2}}) \sqrt{\delta} \sin(\sqrt{\delta} t)  \right.
 \nonumber\\ &&
 \left. \mp \Gamma_{S_{2/1}} (\varepsilon_{\sigma}+\varepsilon_{-\sigma}) \sin\phi \left(1-\cos(\sqrt{\delta} t) \right) \right]
 \nonumber\\
 &+& {{\Gamma_N \Gamma_{S_{1/2}}} \over 4 \pi } \Re\left\{ \left( \Gamma_{S_{1/2}}+\Gamma_{S_{2/1}} e^{\pm i\phi} \right)  \Phi_{\sigma} \right\}
\nonumber \, .
\end{eqnarray}
Using the relation for the induced pairing, Eq.~\ref{eq13}, the above current can be recast as $j_{S_{j}
\sigma}(t) =\Im \left(\Gamma_{S_j} e^{i\varphi_j}  \langle c_{\downarrow}(t)c_{\uparrow}(t) \rangle^* \right)$.
Assuming that $\langle c_{\downarrow}(t)c_{\uparrow}(t) \rangle  = | \langle c_{\downarrow}(t)c_{\uparrow}(t) \rangle |
e^{i\varphi_d}$, where $\varphi_d$ is the argument (phase) of $\langle c_{\downarrow}(t)c_{\uparrow}(t) \rangle$, we
obtain (e.g. \cite{070}):
\begin{eqnarray}\label{eq13cc}
j_{S_{j }\sigma}(t)&=& \Gamma_{S_j} \, |\langle c_{\downarrow}(t)c_{\uparrow}(t) \rangle | \, \sin(\varphi_j-\varphi_d)
\,,
\end{eqnarray}
where $j=1,2$. Inspecting (\ref{eq13cc}) we conclude that the currents flowing between the QD and a given
superconducting lead does not depend on spin, $j_{S_{1 \sigma}}(t)=j_{S_{1 -\sigma}}(t)$, irrespective of 
the spin dependent QD energy levels. This is a consequence of the fact that the QD can exchange charge 
with the superconducting leads only vi pairs of opposite spin electrons.

This formula simplifies for the case $\Gamma_{S_1}=\Gamma_{S_2}\equiv \Gamma_S$ and
$\varepsilon_{\uparrow}+\varepsilon_{\downarrow}=0$, when we obtain
\begin{eqnarray}\label{eq13ff}
j_{S_1\sigma}(t)&=& {\Gamma_{S} \over 2 }  e^{-\Gamma_N t}  \left[1-n_{\sigma}(0)-n_{-\sigma}(0)\right] \,
 \\
 && \cdot \cos(\phi/2) \, \sin(2\Gamma_S |\cos(\phi/2)| t)
 \nonumber\\
&+& {\Gamma_N \Gamma^2_{S} \over 2 \pi }   \cos^2(\phi/2) \, \Re\left\{\Phi_{\sigma}\right\} -{\Gamma_N \Gamma^2_{S}
\over 4 \pi } \sin\phi \, \Im\left\{\Phi_{\sigma}\right\}  \,. \nonumber
\end{eqnarray}
For  $\mu_N=0$ the real part of $\Phi_{\sigma}$, $\Re\left\{\Phi_{\sigma}\right\}$, vanishes\cite{081} and
in such case  for $\phi=\pi$  and identical couplings to both superconducting leads the currents $j_{S_j\sigma}(t)$
vanish.

Under non-equilibrium conditions $\mu_N\neq 0$, for symmetric couplings and for $\varepsilon_{\uparrow}
=-\varepsilon_{\downarrow}$), the asymptotic (for $t\rightarrow \infty$) value of the superconducting 
current can be expressed as
\begin{eqnarray}\label{eq13gg}
j_{S_1\sigma} &=& {\Gamma^2_N \Gamma^2_{S} \over 4 \pi }   \left\{
 \int    {(1-f_N(\varepsilon)) d\varepsilon \over \left[(\Gamma_N^2/4+\varepsilon_{+-}^2)  \right] \left[(\Gamma_N^2/4+\varepsilon_{--}^2)  \right]} \right.
 \nonumber\\
 &-& \left. \int    {f_N(\varepsilon) d\varepsilon \over \left[(\Gamma_N^2/4+\varepsilon_{-+}^2)  \right]
\left[(\Gamma_N^2/4+\varepsilon_{++}^2)  \right]   } \right\}    \cos^2\left(\frac{\phi}{2}\right)
 \nonumber\\
&-& {\Gamma_N \Gamma^2_{S} \over 4 \pi }   \left\{ \int
{(1-f_N(\varepsilon))(\varepsilon+\varepsilon_{\uparrow} ) d\varepsilon \over \left[(\Gamma_N^2/4+\varepsilon_{+-}^2)  \right]
\left[(\Gamma_N^2/4+\varepsilon_{--}^2)  \right] } \right.
 \\
 &-& \left.  \int
{f_N(\varepsilon)(\varepsilon-\varepsilon_{\uparrow} ) d\varepsilon \over \left[(\Gamma_N^2/4+\varepsilon_{-+}^2)  \right]
\left[(\Gamma_N^2/4+\varepsilon_{++}^2)  \right] } \right\} \sin\phi \nonumber
\end{eqnarray}
%
where $\varepsilon_{\alpha \beta}=\varepsilon + E_{\alpha \beta}$ and $E_{\alpha \beta}$ denote 
quasiparticle energies of the proximitized QD. Notice, that the first term in the above formula 
vanishes for zero temperature and $\mu_N=0$. In this case the superconducting current  can be 
rewritten to the following (Josephson-like) formula
\begin{eqnarray}\label{eq13gg2}
j_{S_1\sigma} &=& {\Gamma_S  \over 4 \pi } \frac{ \sin \phi}{|\cos\left(\frac{\phi}{2}\right) |}
 \\
 && \left[ \arctan \frac{ {\varepsilon_{\uparrow}^2+\frac{\Gamma^2_N}{4} }-\Gamma_S^2 \cos^2\left(\frac{\phi}{2}\right)}{\Gamma_S \Gamma_N |\cos\left(\frac{\phi}{2}\right)| }
-\frac{\pi}{2} \right] \,.
  \nonumber
\end{eqnarray}
%
Let us remark that the formula for the current, Eq.~\ref{eq13ff}, can be used to determine the coupling value
$\Gamma_S$. As the time-oscillations  are described by the first term of Eq.~\ref{eq13ff} then for a given $\phi$ the
oscillating part of $j_{S_j \sigma}(t)$ is proportional to $\sin \left( 2\Gamma_S |\cos(\phi/2) | t \right)$. The 
period of these oscillations $T=\frac{\pi}{\Gamma_S |\cos(\phi/2) | }$ for the system characterized by a sufficiently
small $\Gamma_S$ and $\phi \simeq \pi$ should be experimentally detectable.

Lower panel in Fig.~\ref{fig02} presents the current $j_{S_1\uparrow}(t)$ as a function of $\phi$ for
$n_{\uparrow}(0)=n_{\downarrow}(0)=0$. The current oscillates with a damping amplitude and for large time it tends to a
non-zero asymptotic value given in Eq.~\ref{eq13gg2}. The asymptotic value of the current does not depend on the
initial QD occupancies, see Eq.~\ref{eq13dd2}. However, the transient currents are different for the QD initial
occupancies,  $n_{\sigma}(0)=(0,0)$, $(1,1)$  and for $n_{\sigma}(0)=(0,1)$, $(1,0)$. In the first case the current
indicates rather rich time-dependent structure before it attains the asymptotic value. This is a consequence of the
Rabi-like oscillations (damped via $e^{-\Gamma_N t}$ due to the coupling with normal lead) between the empty and double
occupied QD configurations and is described by the first term of Eq.~\ref{eq13dd2} which depends on the factor
$(1-n_{\sigma}(0)-n_{-\sigma}(0))$. This factor disappears for the initial occupancies $n_{\sigma}(0)=(0,1)$ or
$(1,0)$ and all time-dependence of $j_{S_i \sigma}(t)$ is described by the last term of Eq.~\ref{eq13dd2}. This term,
however, in contrast to the former case does not introduce any visible oscillations for small $\Gamma_N$ but enters
the formulas for $j_{S_i\sigma}$, irrespective of the initial conditions. From Fig.~\ref{fig02}
we can learn, that at short time after the quench the current is symmetric with respect to 
$\phi=\pi$. This symmetry, however, is quickly lost and in the long time scale.

\begin{figure}
\centering
\includegraphics[width=80mm]{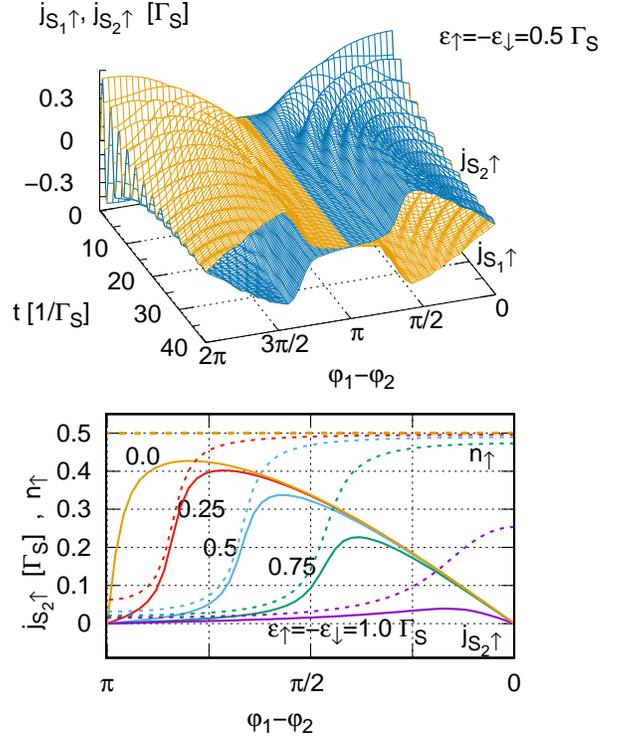}
\caption{\label{fig_zee2} Time dependent currents flowing between the QD and the superconducting  leads,
$j_{S_1\uparrow}(t)$, $j_{S_2\uparrow}(t)$ as a function of the phase difference $\varphi_1-\varphi_2$ in the presence
of the Zeeman splitting $\varepsilon_{\uparrow}=-\varepsilon_{\downarrow}=0.5$ (upper panel). In the bottom panel the
asymptotic spin up currents (solid curves) and corresponding QD occupancies (broken curves) obtained for $t=\infty$ are
shown for the Zeeman splitting: $\varepsilon_{\uparrow}=-\varepsilon_{\downarrow}=0, 0.25, 0.5, 0.75$ and 1.0. The
parameters are: $\Gamma_{S_1}=\Gamma_{S_2}=\Gamma_S$, $\Gamma_N=0.2$ and $n_{\sigma}(0)=0$.  }
\end{figure}
%

In Fig.~\ref{fig_zee2} we present time dependent currents $j_{S_1 \uparrow}$ and $j_{S_2\uparrow}$ 
vs. the phase difference $\phi$ for the finite Zeeman splitting of energy levels, 
$\varepsilon_{\uparrow}=-\varepsilon_{\downarrow}=0.5 \Gamma_S$. Both currents oscillate 
with the period, dependent on the phase difference $\phi$. As before, this period increases 
with $\phi$ and for $\phi=\pi$ the currents do not flow in the system.  Comparing such 
$\phi$-dependence of the currents with those presented in Fig.~\ref{fig02} (lower panel) 
 for $\varepsilon_{\sigma}=0$ we observe a different behavior, especially at asymptotic
large time. In presence of the Zeeman splitting the asymptotic currents almost vanish 
for some $\phi$ interval around $\phi=\pi$.  To study this effect in more detail we  
show in Fig.~\ref{fig_zee2}, bottom panel, the superconducting currents for several 
values of the Zeeman splittings (solid lines for $\varepsilon_{\uparrow}=-\varepsilon_{\downarrow}=0,
0.25, 0.5, 0.75$ and $1.0$ (in units of  $\Gamma_S$). As one can see, in absence of 
the Zeeman splitting the current does not flow only for $\phi=0, \pi$. In presence 
of the Zeeman term the zero-value superconducting current interval of $\phi$ 
increases, but at the same time the maximal values of the currents diminish. 
For $\varepsilon_{\uparrow}=-\varepsilon_{\downarrow} \gg 1$ the superconducting 
currents do not flow. The corresponding asymptotic occupancies of the QD, 
$n_{\uparrow}(\phi,t=\infty)$ are shown in Fig.~\ref{fig_zee2}, bottom panel 
(broken lines). One can notice, that the occupancies decrease monotonically 
with $\phi$ and remain very low for the zero-current interval of $\phi$. 
The changes of the QD occupancies  in presence of the Zeeman splitting  
reflect phasal-dependence of the superconducting currents. These changes are 
related to $0-\pi$ transition and will be discussed in the next paragraph 
(compare $\phi$-dependence of $n_{\uparrow}$ for $\varepsilon_{\uparrow}
=-\varepsilon_{\downarrow}=0.5$ and $\Gamma_N=0.1$ shown in the lower 
panel of Fig.~\ref{fig03new}).

%
\begin{figure}
\centering
\includegraphics[width=65mm]{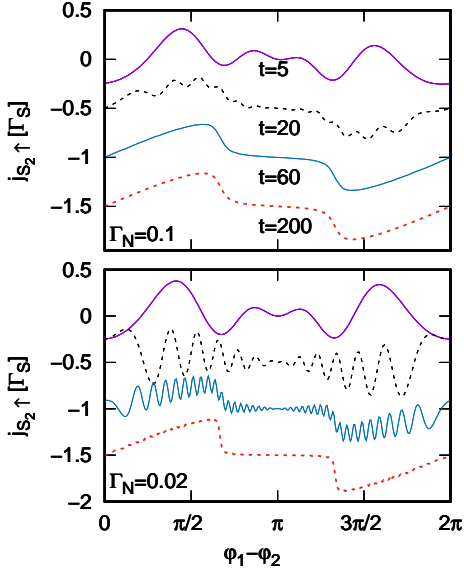}
\caption{\label{fig_zee5}  The current  $j_{S_2\uparrow}(t)$ as a  function of the phase difference
$\varphi_1-\varphi_2$ for different times: $t=5, 20, 60$ and 200 u.t. (from upper to bottom curves in both panels,
respectively). The upper (bottom) panel corresponds to $\Gamma_N=0.1 (0.02)$  and the Zeeman splitting is
$\varepsilon_{\uparrow}=-\varepsilon_{\downarrow}=0.5$, $\Gamma_{S_1}=\Gamma_{S_2}=\Gamma_S$, and $n_{\sigma}(0)=0$.
The curves for $t=20, 60, 200$ are shifted down by 0.5, 1.0, 1.5, respectively, for better visualization.}
\end{figure}
%
In Fig.~\ref{fig_zee5} we analyze the time dependence of $j_{S_2\sigma}(t)$ for some selected values of time $t$, 
starting from the quench $t=0$ until nearly the asymptotically large times. In the lower panel, $\Gamma_N=0.02$, 
the $\phi$-dependence of the current demonstrates abrupt changing of the current value at points corresponding to
$E_{++}=E_{--}$ (see upper panel in Fig.~\ref{fig03new}). These jumps of the current are clearly visible for large
times. However, for larger $\Gamma_N$, e.g. for $\Gamma_N=0.1$ (upper panel, Fig.~\ref{fig_zee5}) the $\phi$-dependence
of the current even for asymptotically large times does not show such sharp changes. Notice, that the time at which
the current achieves constant (in time) values is much shorter in comparison to the case of $\Gamma_N=0.02$. In both
regimes of $\Gamma_N$ we can estimate this time as $\frac{4}{\Gamma_N}$ (compare the results for $\phi$-dependence of
$n_{\sigma}(t)$). For small time  $0-\pi$  transition is not visible but for larger time it becomes evident in spite of the
oscillations. Such transition is very well visible in the asymptotics, where the oscillations vanish. For larger
$\Gamma_N$ the current tends to its asymptotic value (without time-oscillations) in much shorter time than for smaller
$\Gamma_N$, due to the damping factor $e^{-\Gamma_N t}$ [see the first term in Eq.~\ref{eq13ff}].

Let us consider the simple case of the QD coupled solely to superconducting leads, assuming
$\Gamma_{S_1}=\Gamma_{S_2}=\Gamma_{S}$, $\varepsilon_{\uparrow}=-\varepsilon_{\downarrow}$ and
$n_{\uparrow}(0)=n_{\downarrow}(0)=0$. In this case 
\begin{eqnarray}\label{eq13hh}
n_{\sigma}(t)&=&  \sin^2(\Gamma_S \cos(\phi/2)t) \,,
\\ \label{eq13ii}
j_{S_{1/2}\sigma}(t)&=& {\Gamma_{S} \over 2 }  \cos(\phi/2) \sin(2\Gamma_S |\cos(\phi/2)| t) \,.
\end{eqnarray}
The QD occupancy and the current do not depend on spin and, in addition, both superconducting currents,
$j_{S_{1/2}\sigma}(t)$, are exactly identical. Note, however, that for
$\varepsilon_{\uparrow}+\varepsilon_{\downarrow}\neq 0$ these currents differ one from another, see Eq.~\ref{eq13dd2},
and their difference equals $\Gamma_S^2 \sin\phi \, (1-\cos(\sqrt{\delta} \, t))
(\varepsilon_{\uparrow}+\varepsilon_{\downarrow})/\delta$. The current $j_{S_{j}\sigma}$ vanishes for
$\phi=\pi$ and $\Gamma_{S_1}=\Gamma_{S_2}$. For different couplings, $\Gamma_{S_1}\neq \Gamma_{S_2}$, 
 the current does not vanish, even for $\phi=\pi$. For instance $j_{S_1\sigma}$ in this case 
(for $\varepsilon_{\sigma}=0$) is found to be $j_{S_1\sigma}(t)= {\Gamma_{S_1} \over 2 } 
\sin\left[ (\Gamma_{S_1}-\Gamma_{S_2})\, t\right]$.

It would be interesting to consider the transition from the permanently oscillating superconducting currents in the
system of the QD placed only between  two superconducting leads ($\Gamma_N=0$) to finite constant asymptotic values of
such currents in the presence of the third normal lead ($\Gamma_N \neq 0$), see e.g. bottom panel in
Fig.~\ref{fig_zee2}. From Eq.~\ref{eq13ff} we see, that for $\Gamma_N \neq 0$ the current consists of two parts. The
first one corresponds to the transient oscillations damped by the factor $e^{-\Gamma_N t}$, whereas the second one is
described by the imaginary part of $\Phi_{\sigma}$. This part of the current slowly evolves in time to some
non-zero asymptotic value given in Eq.~\ref{eq13gg2}. Asymptotic value of this current vanishes with
decreasing $\Gamma_N$, therefore the oscillating part of is damped less and less effectively, 
and simultaneously the imaginary part of $\Phi_{\sigma}$ vanishes thereby the current oscillates 
with constant amplitude $\frac{\Gamma_S}{2} \cos(\phi/2)$, Eq.~\ref{eq13gg2}.

\section{\label{sec60}Differential subgap conductance}

The last part of our studies is devoted to the subgap time-dependent Andreev conductance 
$G_{\sigma}(\mu,t)={\partial \over \partial \mu} j_{N\sigma}(t)$, expressing it in units 
of ${4 e^2 \over h}$. We investigate this quantity as a function of the bias voltage 
($\mu=\mu_N$) applied to the normal lead. Using the expressions for the current and 
QD charge, Eqs.~\ref{eq08} and \ref{eq10bb}, we obtain at zero temperature
\begin{eqnarray}\label{eq14}
G_{\sigma}(\mu, t)&=&  \Re \left[ {\Gamma_{N}} e^{-i \mu t}  \mathcal{L}^{-1}\left\{
{s+i\varepsilon_{-\sigma}+\Gamma_N/2 \over (s-s_1)(s-s_2)(s-i\mu)}\right\}(t)     \right]
 \nonumber\\
 &+&{ \Gamma_N^2 \Gamma_{12} \over 8} \mathcal{L}^{-1}\left\{ {1 \over (s-s_1)(s-s_2)(s+i\mu)}\right\}(t)
 \nonumber\\
 && \mathcal{L}^{-1}\left\{ {1 \over (s-s_3)(s-s_4)(s-i\mu)}\right\}(t)
  \\
 &-&  {\Gamma_{N}^2 \over 2} \mathcal{L}^{-1}\left\{ {s+i\varepsilon_{-\sigma}+\Gamma_N/2 \over (s-s_1)(s-s_2)(s-i\mu)
 }\right\}(t)
  \nonumber\\
 && \mathcal{L}^{-1}\left\{ {s-i\varepsilon_{-\sigma}+\Gamma_N/2 \over (s-s_3)(s-s_4)(s+i\mu) }\right\}(t) \nonumber
\end{eqnarray}
where for $\sigma=\downarrow$ the replacement $(s_1,s_2) \rightarrow (s_3,s_3)$ should be made. Notice, that for
$\varepsilon_{\uparrow}=\varepsilon_{\downarrow}$ the conductance is spin independent
($G_{\uparrow}=G_{\downarrow}=G$).
%
\begin{figure}
\centering
\includegraphics[width=80mm]{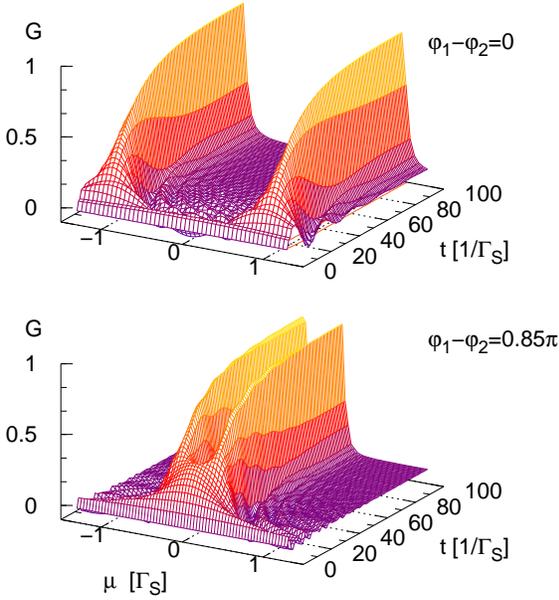}
\caption{\label{fig04} Time-dependent Andreev conductance (in ${4 e^2 / h} $ units) as a function of the bias voltage
$\mu$ for the phase difference  $\phi=\varphi_1-\varphi_2=0$ (upper panel) and for $\phi=0.85\pi$ (bottom panel). The
other system parameters are the same as in Fig.~\ref{fig01}.}
\end{figure}
%
In Fig.~\ref{fig04} we plot the time-dependent conductance $G_{\sigma}(\mu,t)=G$ as a function of $\mu$ for different
phase difference between the superconducting leads, i.e.\ for $\phi=0$ (upper panel) and for $\phi=0.85\pi$ (bottom
panel), in the presence of weakly coupled normal electrode, $\Gamma_N =0.1\Gamma_S$
($\Gamma_{S_1}=\Gamma_{S_2}=\Gamma_S=1$) and $\varepsilon_{\sigma}=0$. The process of forming the Andreev subgap
states is clearly visible. We observe that for $\phi=0$ in the limit of large time the conductance is characterized by
two well pronounced maxima appearing at $\mu\simeq \pm \Gamma_S$ whose half-widths gradually shrink in time.
These maxima appear after some time-interval after abrupt switching of the QD-leads couplings (we denote such
time-scale by $\tau_1$ see Fig.~\ref{fig04b}). This characteristic time is needed to build up two distinct 
maxima of $G$ is longer depends on the phase difference $\phi$ -- compare the upper and bottom panels in 
Fig.~\ref{fig04}. Time-evolution of such quasiparticle peaks  allows us to estimate how fast the Andreev 
quasipatricles appear in the system and thus it is desirable to study this process in more detail.

By inspecting $G_{\sigma}(\mu,t)$  in Fig.~\ref{fig04} we observe, that up to some specific time,
$\tau_1$, a broad one-peaked structure of $G$ is present. Then, the conductance rapidly transforms 
in time into two-peak structure. The position of each quasiparticle peak evolves in time to its  
steady limit value (that time is called $\tau_2$) and finally the width and height of peaks are 
established after the time $\tau_f$ (see Fig.~\ref{fig01}, bottom panel).
%
\begin{figure}
\centering
\includegraphics[width=80mm]{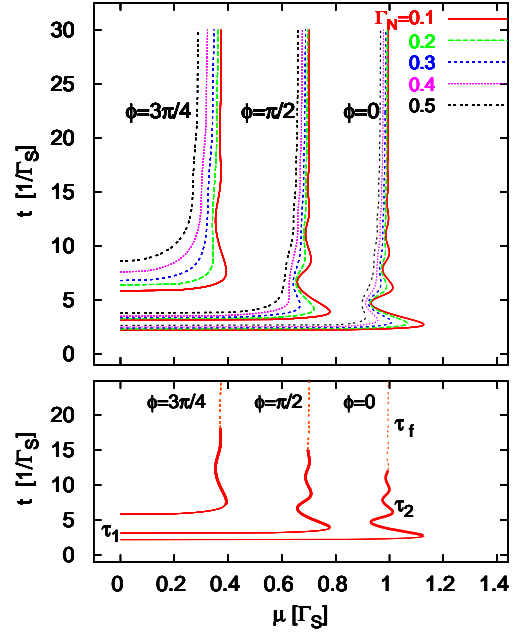}
\caption{\label{fig04b} Positions of the quasiparticle maxima  vs. time and $\mu$ appearing in the differential
conductance $G_{\sigma}(\mu,t)$ for different $\Gamma_N$ indicated in the legend and for superconducting phase
difference $\phi=0, \pi/2$ and $3\pi/4$, respectively (upper panel). The bottom panel shows the result for
$\Gamma_N=0.1$  where different time scales, $\tau_1$, $\tau_2$ and $\tau_3$ are
indicated. For negative values of $\mu$ the results are symmetrical. The QD energy levels are: $\varepsilon_{\sigma}=0$
and $\Gamma_{S_1}=\Gamma_{S_2}=\Gamma_S$. }
\end{figure}
%
In Fig.~\ref{fig04b} we display  position of the quasiparticle peaks maxima vs. time and $\mu$ for different values
of $\Gamma_N$ and $\phi$ indicated in the legend (upper panel). As one can see, the moment of appearance of two-peak
structure, $\tau_1$, depends on both $\phi$ and $\Gamma_N$. However, for $\phi=0$ this time only slightly depends on
$\Gamma_N$. With increasing $\phi$ it  increases with remarkable dependence  on $\Gamma_N$ (for a given $\phi$ it
increases with $\Gamma_N$). The time scale for appearance of the two-peak structure is very small and for $\phi=0$ it
equals approximately 2.5 u.t.,  for $\phi=\pi/2$ it changes from $\sim 3$ u.t. for $\Gamma_N=0.1$ up to $\sim 4$ u.t.
for $\Gamma_N=0.5$, and for $\phi=3\pi/4$  it changes from  $\sim 6$ u.t. for $\Gamma_N=0.1$ up to $\sim 8.5$ u.t. for
$\Gamma_N=0.5$, respectively (see upper panel, Fig.~\ref{fig04b}).
Positions of the  maxima versus $\mu$ evolve in time during approximately $\tau_2\simeq 12$ u.t. (the bold
parts of lines in the bottom panel) and attain their steady-state values. Note, that the asymptotic quasiparticle peaks
heights and widths are achieved  with the envelope function $1-\exp(-t/\tau_f)$, where $\tau_f=\frac{2}{\Gamma_N}$ can be
deduced from the explicit expression for $G_{\sigma}(\mu,t)$ in which the long living terms proportional to
$\exp(-\Gamma_N t/2)$ are present.

Let us consider a few special cases, for which the simpler analytical formulas can be given. In the limit
$\phi=\pi$, $\varepsilon_{\sigma}=0$ and $\Gamma_{S_1}=\Gamma_{S_2}$ the conductance takes the form (here
$G_{\uparrow}=G_{\downarrow}=G$):
\begin{eqnarray}\label{eq15}
G(\mu, t)&=&  {\Gamma_{N} \over  \left({\Gamma_{N}^2 \over 4} +\mu^2 \right)}
 \left[{\Gamma_{N} \over 2} + e^{-\Gamma_N t/ 2} \cdot \right.
 \\ &&
 \left. \left( {\Gamma_{N} \over 2} \cos(\mu t)- \Gamma_N \cosh\left({\Gamma_{N} t\over 2} \right)+\mu \sin(\mu t) \right)
\right] \nonumber \,.
\end{eqnarray}
In this case the zero bias conductance reads $G(\mu=0, t)=  2  \left[1+ e^{-\Gamma_N {t/ 2}} \left(1-2
\cosh\left({\Gamma_{N} t / 2} \right) \right) \right]$ and for $t={2\ln 2 /\Gamma_N}$ it reaches 
the optimal value equal to $0.5$ and it vanishes for $t\rightarrow\infty$.

As differential conductance depends on the couplings $\Gamma_{S_1}$, $\Gamma_{S_2}$ and $\phi$ 
only through $\Gamma_{12}$ so different choices of these parameters can lead to the same values 
of $G_{\sigma}$. Note, that vanishing conductance $G_{\sigma}(\mu, t=\infty)$ for $\phi=\pi$ 
is obtained, assuming the symmetric couplings $\Gamma_{S_1}=\Gamma_{S_2}$. For
$\Gamma_{S_1}\neq \Gamma_{S_2}$ even for $\phi=\pi$ the conductance looks quite different. 
Assuming, e.g. $\frac{\Gamma_{S_2}}{\Gamma_{S_1}}=k$ we obtain $\Gamma_{12}=
\Gamma_{S_1}^2(1+k^2+2k\cos\phi)$, than for $k\neq 1$ and $\phi=\pi$ one has 
$\Gamma_{12}=\Gamma_{S_1}^2(1-k)^2$. The same value of $\Gamma_{12}$ one can 
obtain for $k=1$ and $\phi=arccos\left(\frac{(1-k)^2}{2}-1 \right)$. 
The conductance $G_{\sigma}(\mu, t)$ shown in Fig.~\ref{fig04}
for $\phi=0.85\pi$ and $\Gamma_{S_1}=\Gamma_{S_2}=1$ is identical with that one 
calculated for $\frac{\Gamma_{S_2}}{\Gamma_{S_1}}=\frac{1}{2}$ and 
$\phi=arccos\left(-\frac{7}{8} \right)$. It means that asymmetry in
the couplings to superconducting leads $\Gamma_{S_1}, \Gamma_{S_2}$ 
can be effectively captured by the phase difference parameter,  $\phi$. 
This conclusion refers also to the QD occupancy and the current flowing 
between the QD and the normal lead.  Since explicit expression for 
$G_{\sigma}(\mu, t)$ is rather lengthly, we skip it here and present 
only its asymptotic form  ($t\rightarrow \infty$) for $\Gamma_{S_1}
=\Gamma_{S_2}=\Gamma_S$, $\varepsilon_{\uparrow}=-\varepsilon_{\downarrow}$ 
($G_{\uparrow}=G_{\downarrow}=G$)
\begin{eqnarray}\label{eq14b}
G(\mu)&=&  \frac{\Gamma_{N}^2 \Gamma_{S}^2}{2} \cos^2\left(\frac{\phi}{2}\right) \left\{ \frac{1}{
 \left( \frac{\Gamma_N^2}{4}+\mu_{-+}^2 \right) \left( \frac{\Gamma_N^2}{4}+\mu_{++}^2 \right) }   \right.
 \nonumber\\
 &+& \left.  \frac{1}{
 \left( \frac{\Gamma_N^2}{4}+\mu_{+-}^2 \right) \left( \frac{\Gamma_N^2}{4}+\mu_{--}^2 \right) }  \right\} \,,
\end{eqnarray}
where $\mu_{\alpha \beta}=\mu+E_{\alpha \beta}$. For $\Gamma_N \ll \Gamma_S$ the asymptotic
conductance has four maxima placed at $\mu \simeq \pm \varepsilon_{\uparrow}\pm \Gamma_S 
|\cos\left( \frac{\phi}{2}\right)|$ or equivalently at $\mu=E_{++}, E_{+-}, E_{-+}$ and $E_{--}$, 
respectively. Note that the asymptotic conductivity $G(\mu)$ does not depend on spin but 
in general $G_{\sigma}(\mu, t)$ it can be spin-dependent.

\begin{figure}
\centering
\includegraphics[width=80mm]{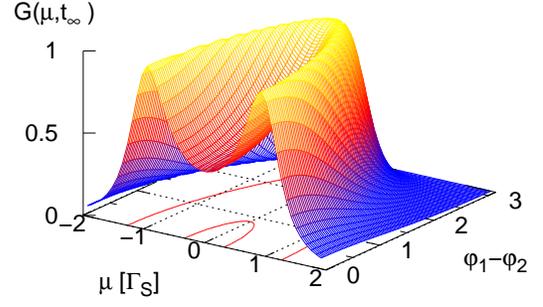}
\caption{\label{fig05} The asymptotic conductance obtained for $t\rightarrow \infty$ as a function of $\mu$ and the
phase difference $\phi=\varphi_1-\varphi_2$. The contour lines correspond to $G=0.5$ and
$\Gamma_{S1}=\Gamma_{S2}=\Gamma_S$, $\Gamma_N=0.75$. }
\end{figure}

It is also interesting to check influence of the superconducting phase difference 
on the the asymptotic Andreev conductance behavior. For arbitrary $\phi \neq \pi$ 
and $\varepsilon_{\sigma}=0$, the asymptotic value of $G(\mu,t)$ can be written 
as follows (for $t=\infty$)
\begin{eqnarray}\label{eq16}
G(\mu)&=&  {\Gamma_{N}^2 \Gamma_{12} \over  4 \left[ {\Gamma_{N}^2 \over 4} +\left({\sqrt{\Gamma_{12}} \over 2}- \mu
\right)^2  \right] \left[ {\Gamma_{N}^2 \over 4} +\left({\sqrt{\Gamma_{12}}\over 2 }+ \mu \right)^2  \right] } \,.
\nonumber\\
\end{eqnarray}
Fig.~\ref{fig05} presents the asymptotic conductance, $G(\mu, t=\infty)$, as a function of the bias voltage
$\mu$, and the phase difference $\phi$. As one can see for $\phi=0$ two distinct maxima of $G$ are visible 
(cf. Fig.~\ref{fig04} for  $t=100$).
%
%
For nonzero ${\phi}$, which satisfies the condition  $\cos({\phi})> {\Gamma_N^2-\Gamma_{S1}^2-\Gamma_{S2}^2 \over
2\Gamma_{S1}\Gamma_{S2}}$,  these two maxima appear at points $\mu=\pm \sqrt{{\Gamma_{12}\over 4}-{\Gamma_N^2\over
4}}$. In the opposite case, there is only one maximum at $\mu=0$ whose height is reduced to zero value with $\phi
\rightarrow \pi$. In consequence, for $\phi=\pi$ and $t=\infty$ the conductance vanishes for all $\mu$. Note that, 
for the QD coupled only to one superconducting and one normal electrode, the zero-bias conductance is invariant 
under the replacement $\Gamma_N\leftrightarrow\Gamma_S$, \cite{071}. However, in our system with two 
superconducting leads this conclusion is no longer valid, even for the symmetric couplings case, 
$\Gamma_{S_1}=\Gamma_{S_2}$. Such property is achieved only for $\phi={2\pi \over 3}$.

%
\begin{figure}
\centering
\includegraphics[width=65mm]{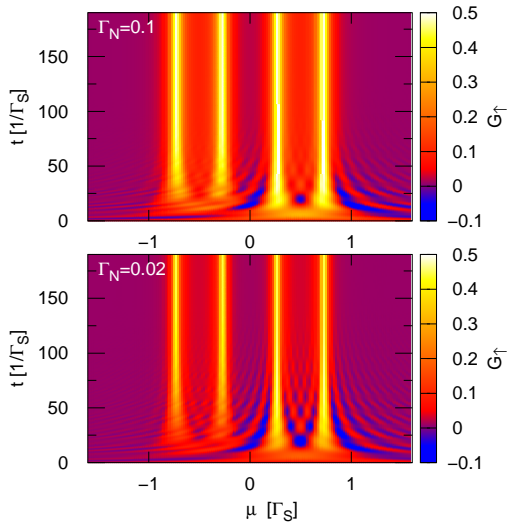}
\caption{\label{fig11} Time-dependent Andreev conductance $G_{\uparrow}$ (in ${4 e^2 / h} $ units) as a function of the
bias voltage $\mu$ for the phase difference  $\phi=\varphi_1-\varphi_2=0.85\pi$ and the Zeeman splitting
$\varepsilon_{\uparrow}=-\varepsilon_{\downarrow}=0.5$. The upper (bottom)  panel corresponds to $\Gamma_N=0.1$
($\Gamma_N=0.02$) and $\Gamma_{S1}=\Gamma_{S2}=\Gamma_S$.}
\end{figure}
\begin{figure}
\centering
\includegraphics[width=75mm]{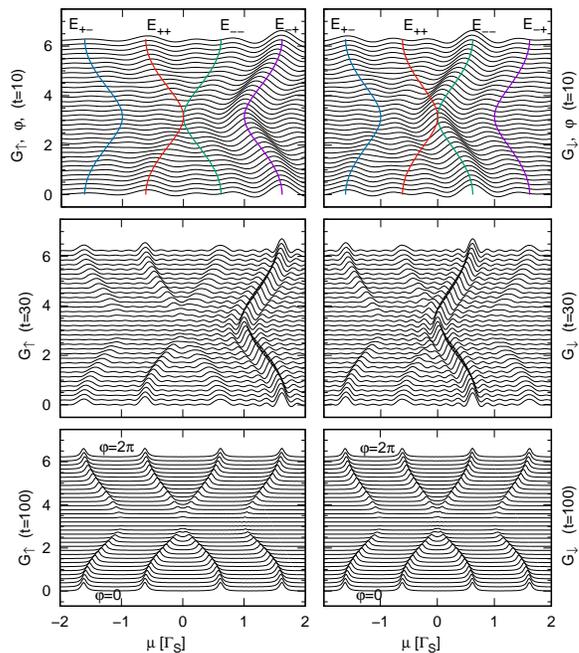}
\caption{\label{fig12} Phase-dependent Andreev conductance $G_{\uparrow}$ (left panels) and $G_{\downarrow}$ (right panels)
(in ${4 e^2 / h} $ units) as a function of the bias voltage $\mu$ for $t=10, 30$ and 100 u.t. (upper, middle and bottom panels,
respectively) after the quench. Other parameters: $\varepsilon_{\uparrow}=1$, $\varepsilon_{\downarrow}=0$,  $\Gamma_N=0.1$,
$\Gamma_{S1}=\Gamma_{S2}=\Gamma_S$. The lower curves in all panels correspond to $\varphi=0$ and each next upper curve is shifted
up by $\frac{2\pi}{33}$, so the upper curves correspond to  $\varphi=2\pi$. $E_{+-}, E_{++}, E_{--}, E_{-+}$ are represented by corresponding
solid lines in the upper panels:  they show the localization of the ABS (for $\Gamma_N=0$) on the $(\mu, \varphi)$ plane.
}
\end{figure}

%
In the last part of our studies  we discuss the time-evolution of the ABS for nonzero splitting of the QD energy
levels. In the first case  we consider the symmetric splitting around the zero energy (Fig.~\ref{fig11},
$\varepsilon_{\uparrow}=-\varepsilon_{\downarrow}=0.5$ for $\phi=0.85\pi$) and in the second case the splitting 
is  symmetric but around the nonzero energy value equal 0.5 (Fig.~\ref{fig12}, $\varepsilon_{\uparrow}=1$, $\varepsilon_{\downarrow}=0$ for some specific values of time after the quench). In Fig.~\ref{fig11} we analyze the
approach to equilibrium of $G_{\uparrow}(\mu,t)$ for two values of $\Gamma_N$, $\Gamma_N=0.1 (0.02)$ upper (bottom)
panel. We show only $G_{\uparrow}(\mu,t)$ as $G_{\downarrow}(\mu,t)$ is symmetric (with respect to $\mu=0$) 
in relation to $G_{\uparrow}$.  The maxima of $G_{\uparrow}$ for large time correspond to $E_{-+}$, $E_{++}$, 
$E_{--}$ and $E_{+-}$ ABS states (on the negative side of $\mu$-axis). It is interesting that the time 
evolution of $E_{-+}, E_{++}$ ABS is different from the evolution of $E_{--}$ and $E_{+-}$, respectively. 
The stationary values of the conductance peaks corresponding to $G_{\uparrow}$ and $G_{\downarrow}$ are 
all the same (according to Eq.~\ref{eq14b}) but the ABS $E_{-+}$ and $E_{++}$ begin to appear later 
than $E_{--}$ and $E_{+-}$. For $\Gamma_N=0.1$ (0.02) this delay time can be approximately estimated 
as 30 (60) u.t. For stronger coupling $\Gamma_N$ (upper panel) the ABS peaks are wider in comparison 
to the case of weakly coupled normal electrode (bottom panel) and appear earlier than for smaller $\Gamma_N$.
%
%
In Fig.~\ref{fig12} we show the phasal-dependence of $G_{\uparrow}$ and $G_{\downarrow}$ calculated 
for small time, $t=10$ u.t. (upper panels), for $t=30$ u.t. (middle panels) and for long time, 
$t=100$ u.t. (bottom panels) at which the conductance attains the stationary values.
In addition, in the upper panels the curves representing the localization of the ABS states 
on the   $(\mu, \varphi)$ plane are depicted. We observe essential difference with strong 
asymmetry between  $G_{\uparrow}$ and $G_{\downarrow}$ at short period of time after the quench.
The time evolution of  $G_{\uparrow}$ ($G_{\downarrow}$) is limited to the appearance of 
$E_{-+}$ ($E_{--}$) ABS. Next, for larger time other Andreev states appear but the most 
visible are still the curves corresponding to $E_{-+}$ and $E_{--}$, respectively. Notice,
that for $\varphi=\pi$ and large time the conductance vanishes for both spins (cf. Eq.~\ref{eq14b}) 
as shown in the bottom panels. However, for smaller time after the quench all ABS states 
also vanish for  $\varphi=\pi$ except $E_{-+}$ (for $\sigma=\uparrow$) and $E_{--}$ 
(for $\sigma=\downarrow$). These states vanish only at relatively large times after 
the quench.

\section{\label{sec70}Conclusions  }

We have analyzed theoretically the transient sub-gap quasiparticle and transport 
properties of the QD coupled to one metallic and two superconducting electrodes 
(with large energy gaps), using the equation of motion for the second quantization
operators and determining their Laplace transforms. Response of the system to the 
sudden coupling of its constituent parts and influence of the initial QD 
occupations on the induced electron pairing, the transient currents, and the 
differential conductance have been studied. The analytical formulas for these 
quantities have been derived and for some specific situations and their
mutual relations have been analyzed. We have addressed the equilibrium case 
(with identical chemical potentials of all leads) and investigated 
the conductivity on non-equilibrium (biased) system.

We have shown that formulas for the QD occupation, $n_{\sigma}(t)$, on-dot 
pairing amplitude $\langle c_{\downarrow}(t) c_{\uparrow}(t)\rangle$ and 
charge  current flowing between the QD and superconducting leads, $j_{S_j \sigma}(t)$, 
consist of two parts. The first one depends on the initial QD occupancies, but 
does not depend on the chemical potentials of external reservoirs. This term 
describes the oscillating transient behavior of the considered quantities 
which is damped via $\exp(-\Gamma_N t)$ due to the QD-normal lead coupling. 
It is proportional to the factor $(1-n_{\uparrow}(0)-n_{\downarrow}(0))$ and 
vanishes for some specific initial QD filling. The second part of the considered 
formulas depends mainly on the QD-normal leads coupling and results in monotonic 
time-dependence of the corresponding quantities. In contrast to the first 
part, it appears in all formulas regardless of the initial conditions.

Having analytical expressions for the physical observables we have shown, how 
the amplitude and time period of the transient oscillations depend on the model
parameters $\varepsilon_{\sigma}$, $\Gamma_{S_i}$, $\Gamma_N$ and $\varphi_1$,
$\varphi_2$. We have also presented a reminiscence of the Rabi-type oscillations 
between the empty and doubly-occupied configurations of the proximitized QD.     
We have found that for $\varepsilon_{\sigma}=0$ the asymptotic QD occupancy 
does not depend on the superconducting leads phase difference $\phi$ and 
it tends to half-filling. In presence of the Zeeman splitting $n_{\sigma}(t)$ 
relevantly depends on the phase difference $\phi$, indicating signatures of 
$0-\pi$ transition. Such transition is at smaller time (right after the quench) 
rather not much evident, but it becomes more and more clear at larger times, 
$t\geq \frac{4}{\Gamma_N}$.

Finally, we have analyzed the time-dependent differential conductance as 
a function of the bias voltage between the normal and superconducting
leads and we have inspected its phasal dependence. It has been found, 
that two-peak structure of the conductance (for $\varepsilon_{\sigma}=0$) 
known from the stationary transport properties, does emerge after 
some characteristic time-interval. This time scale increases with
the phase differences $\phi$. We have analyzed the spin-dependent 
conductance considering the Zeeman splitting of the QD levels,
and found different temporal evolution of the corresponding 
Andreev bound states. Ultimately, for asymptotically large times, 
these Andreev peaks become symmetric and spin-independent.
Our theoretical predictions could be verified by the present-day experimental
methods and they could shed light on dynamics of the sub-gap quasiparticle
states.

\section*{Acknowledgements}
This work was supported by National Science Centre (NCN, Poland) under 
the grant UMO-2017/27/B/ST3/01911 (T.D. and R.T.).


\bibliography{biblio}

\end{document}